\documentclass[trackchanges,twocolumn,twocolappendix]{aastex701}
\accepted{August 3, 2026}
\submitjournal{The Astronomical Journal}

\usepackage{multirow}
\usepackage{tablefootnote}
\usepackage{soul}

\begin{document}

\title{Atmospheric Reconnaissance of TRAPPIST-1\,f with JWST NIRISS SOSS: No Evidence for the Transit Light Source Effect}

\author[0000-0003-4676-0622]{Olivia Lim}
\affiliation{Institut Trottier de recherche sur les exoplan\`etes, D\'epartement de Physique, Universit\'e de Montr\'eal, Montr\'eal, Qu\'ebec, Canada}
\email{olivia.lim@umontreal.ca}

\author[0000-0001-5485-4675]{Ren\'{e} Doyon}
\affiliation{Institut Trottier de recherche sur les exoplan\`etes, D\'epartement de Physique, Universit\'e de Montr\'eal, Montr\'eal, Qu\'ebec, Canada}
\email[show]{rene.doyon@umontreal.ca}

\author[0000-0003-4816-3469]{Ryan J. MacDonald}
\altaffiliation{NHFP Sagan Fellow}
\affiliation{Department of Astronomy, University of Michigan, 1085 S. University Ave., Ann Arbor, MI 48109, USA}
\email{ryanjmac@umich.edu}

\author[0000-0003-3506-5667]{\'{E}tienne Artigau}
\affiliation{Institut Trottier de recherche sur les exoplan\`etes, D\'epartement de Physique, Universit\'e de Montr\'eal, Montr\'eal, Qu\'ebec, Canada}
\email{etienne.artigau@umontreal.ca}

\author[0000-0002-3328-1203]{Michael Radica}
\altaffiliation{NSERC Postdoctoral Fellow}
\affiliation{Department of Astronomy \& Astrophysics, University of Chicago, 5640 South Ellis Avenue, Chicago, IL 60637, USA}
\affiliation{Institut Trottier de recherche sur les exoplan\`etes, D\'epartement de Physique, Universit\'e de Montr\'eal, Montr\'eal, Qu\'ebec, Canada}
\email{radicamc@uchicago.edu}

\author[0000-0002-9479-2744]{Mykhaylo Plotnykov}
\affiliation{Department of Physics, University of Toronto, Toronto, ON M5S 3H4, Canada}
\email{mykhaylo.plotnykov@mail.utoronto.ca}

\author[0009-0005-6135-6769]{Alexandrine L'Heureux}
\affiliation{Institut Trottier de recherche sur les exoplan\`etes, D\'epartement de Physique, Universit\'e de Montr\'eal, Montr\'eal, Qu\'ebec, Canada}
\email{alexandrine.lheureux@umontreal.ca}

\author[0000-0002-2875-917X]{Caroline Piaulet-Ghorayeb}
\altaffiliation{E. Margaret Burbidge Postdoctoral Fellow}
\affiliation{Department of Astronomy \& Astrophysics, University of Chicago, 5640 South Ellis Avenue, Chicago, IL 60637, USA}
\affiliation{Institut Trottier de recherche sur les exoplan\`etes, D\'epartement de Physique, Universit\'e de Montr\'eal, Montr\'eal, Qu\'ebec, Canada}
\email{carolinepiaulet@uchicago.edu}

\author[0000-0002-2195-735X]{Louis-Philippe Coulombe}
\affiliation{Institut Trottier de recherche sur les exoplan\`etes, D\'epartement de Physique, Universit\'e de Montr\'eal, Montr\'eal, Qu\'ebec, Canada}
\email{louis-philippe.coulombe@umontreal.ca}

\author[0000-0002-6780-4252]{David Lafreni\`{e}re}
\affiliation{Institut Trottier de recherche sur les exoplan\`etes, D\'epartement de Physique, Universit\'e de Montr\'eal, Montr\'eal, Qu\'ebec, Canada}
\email{david.lafreniere@umontreal.ca}

\author[0000-0002-5967-9631]{Thomas J. Fauchez}
\affiliation{NASA Goddard Space Flight Center, 8800 Greenbelt Road, Greenbelt, MD 20771, USA}
\affiliation{Integrated Space Science and Technology Institute, Department of Physics, American University, Washington, DC 20016, USA}
\affiliation{NASA GSFC Sellers Exoplanet Environments Collaboration, 8800 Greenbelt Road, Greenbelt, MD 20771, USA}
\email{thomas.j.fauchez@nasa.gov}

\author[0000-0003-2260-9856]{Martin Turbet}
\affiliation{Laboratoire de M\'{e}t\'{e}orologie Dynamique/IPSL, CNRS, Sorbonne Universit\'{e}, Ecole Normale Sup\'{e}rieure, PSL Research University, Ecole Polytechnique, 75005 Paris, France}
\affiliation{Laboratoire d'astrophysique de Bordeaux, Univ. Bordeaux, CNRS, B18N, all\'{e}e Geoffroy Saint-Hilaire, 33615 Pessac, France}
\email{martin.turbet@lmd.ipsl.fr}

\author[0000-0003-3993-4030]{Diana Valencia}
\affiliation{Department of Physical \& Environmental Sciences, University of Toronto at Scarborough, Toronto, ON M1C 1A4, Canada}
\affiliation{David A. Dunlap Dept. of Astronomy \& Astrophysics, University of Toronto, 50 St. George Street, Toronto, ON M5S 3H4, Canada}
\email{diana.valencia@utoronto.ca}

\author[0000-0003-0475-9375]{Lo\"{i}c Albert}
\affiliation{Institut Trottier de recherche sur les exoplan\`etes, D\'epartement de Physique, Universit\'e de Montr\'eal, Montr\'eal, Qu\'ebec, Canada}
\email{loic.albert@umontreal.ca}

\author[0000-0001-6362-0571]{Laura Flagg}
\affiliation{Astronomy Department, Cornell University, Ithaca, NY 14853, USA}
\affiliation{Carl Sagan Institute, Cornell University, Ithaca, NY, USA}
\email{laura.flagg1@gmail.com}

\author[0000-0001-5578-1498]{Bj\"{o}rn Benneke}
\affiliation{Institut Trottier de recherche sur les exoplan\`etes, D\'epartement de Physique, Universit\'e de Montr\'eal, Montr\'eal, Qu\'ebec, Canada}
\email{bjorn.benneke@umontreal.ca}

\author[0000-0003-4166-4121]{Neil J. Cook}
\affiliation{Institut Trottier de recherche sur les exoplan\`etes, D\'epartement de Physique, Universit\'e de Montr\'eal, Montr\'eal, Qu\'ebec, Canada}
\email{neil.cook@umontreal.ca}

\author[0000-0001-6809-3520]{Pierre-Alexis Roy}
\affiliation{Institut Trottier de recherche sur les exoplan\`etes, D\'epartement de Physique, Universit\'e de Montr\'eal, Montr\'eal, Qu\'ebec, Canada}
\email{pierre-alexis.roy@umontreal.ca}

\author[0000-0002-0436-1802]{Lisa Kaltenegger}
\affiliation{Carl Sagan Institute, Cornell University, Ithaca, NY, USA}
\affiliation{Cornell Center for Astrophysics and Planetary Science, Cornell University, Ithaca, NY 14853, USA}
\affiliation{Astronomy Department, Cornell University, Ithaca, NY 14853, USA}
\email{lkaltenegger@astro.cornell.edu}

\author[0000-0001-9291-5555]{Charles Cadieux}
\affiliation{Institut Trottier de recherche sur les exoplan\`etes, D\'epartement de Physique, Universit\'e de Montr\'eal, Montr\'eal, Qu\'ebec, Canada}
\email{charles.cadieux@umontreal.ca}

\author[0000-0001-6129-5699]{Nicolas B. Cowan}
\affiliation{Department of Physics and Trottier Space Institute, McGill University, 3600 rue University, H3A 2T8 Montreal QC, Canada}
\email{nicolas.cowan@mcgill.ca}

\author[0000-0003-1462-7739]{Micha\"{e}l Gillon}
\affiliation{Astrobiology Research Unit, Universit\'{e} de Li\`{e}ge, All\'{e}e du 6 Ao\^{u}t 19C, B-4000 Li\`{e}ge, Belgium}
\email{michael.gillon@uliege.be}

\author[0000-0002-1199-9759]{Romain Allart}
\altaffiliation{SNSF Postdoctoral Fellow}
\affiliation{Institut Trottier de recherche sur les exoplan\`etes, D\'epartement de Physique, Universit\'e de Montr\'eal, Montr\'eal, Qu\'ebec, Canada}
\email{romain.allart@umontreal.ca}

\author[0000-0003-4987-6591]{Lisa Dang}
\affiliation{Institut Trottier de recherche sur les exoplan\`etes, D\'epartement de Physique, Universit\'e de Montr\'eal, Montr\'eal, Qu\'ebec, Canada}
\email{lisa.dang@uwaterloo.ca}

\author[0000-0002-6773-459X]{Doug Johnstone}
\affiliation{NRC Herzberg Astronomy and Astrophysics, 5071 West Saanich Rd, Victoria, BC, V9E 2E7, Canada}
\affiliation{Department of Physics and Astronomy, University of Victoria, Victoria, BC, V8P 5C2, Canada}
\email{doug.johnstone@nrc-cnrc.gc.ca}

\author[0000-0002-8573-805X]{Stefan Pelletier}
\affiliation{Observatoire astronomique de l'Universit\'e de Gen\`eve, 51 chemin Pegasi 1290 Versoix, Switzerland}
\email{stefan.pelletier@unige.ch}

\author[0000-0001-9987-467X]{Jared Splinter}
\affiliation{Trottier Space Institute at McGill, 3550 rue University, Montr\'eal, QC H3A 2A7, Canada}
\affiliation{Department of Earth and Planetary Sciences, McGill University, 3450 rue University, Montr\'eal, QC H3A OE8, Canada}
\email{jared.splint@gmail.com}

\author[0000-0003-4844-9838]{Jake Taylor}
\affiliation{Department of Physics, University of Oxford, Parks Rd, Oxford OX1 3PU, UK}
\email{jake.taylor@physics.ox.ac.uk}


\begin{abstract}

In just over three years of operation, JWST has observed all seven planets of the TRAPPIST-1 system. 
The two innermost planets were found to have little to no atmosphere, barring the presence of high-altitude aerosols. 
Here we present the first JWST transit spectra of the habitable-zone exoplanet TRAPPIST-1\,f, which were obtained with NIRISS SOSS over the course of five transits. At least one stellar flare occurred in each visit, but unlike observations of closer-in TRAPPIST-1 planets, no evidence for contamination of the transit spectra from unocculted stellar surface heterogeneities was found. This non-detection does not guarantee the absence of unocculted heterogeneities in all future transit observations of this planet, and it could be explained by the transit chord of TRAPPIST-1\,f having properties similar to the average, out-of-transit, visible stellar hemisphere at the time of observation. 
The transit spectra exhibit slopes ranging from $-365\,$ppm/$\mu$m down to 15\,ppm/$\mu$m, which we attribute to stellar variability, that is, flares and/or smaller-scale events. The visits least affected by flares rule out H$_2$/He-dominated atmospheres with surface pressures higher than about 20\,mbar at 95\% confidence. 
For high-mean-molecular-mass atmospheres, the exact upper limits on surface pressures depend on the reduction pipeline and on the treatment of the residual slopes in the transit spectra. 

\end{abstract}

\keywords{\uat{Exoplanets}{498} --- \uat{Extrasolar rocky planets}{511} --- \uat{M dwarf stars}{982} --- \uat{Stellar activity}{1580} --- \uat{Stellar flares}{1603} --- \uat{Exoplanet atmospheres}{487} --- \uat{Transits}{1711}}


\section{Introduction} \label{sec:intro}

The search for atmospheres on small exoplanets with JWST has already begun with some of the most favorable known targets, including planets in the TRAPPIST-1 system. Host to seven Earth-sized, rocky, transiting planets in a compact orbital configuration \citep{gillon_temperate_2016,gillon_seven_2017,luger_seven-planet_2017,gillon_trappist-1_2020,agol_refining_2021}, the TRAPPIST-1 system was observed both in emission and transmission serving as the subject of several JWST GO and GTO programs in Cycle 1--3. 

Secondary eclipse observations of TRAPPIST-1\,b with JWST \citep[MIRI,][]{bouchet_mid-infrared_2015,wright_mid-infrared_2023} at 12.8 and 15\,$\mu$m rejected several atmospheric scenarios that include CO$_2$ absorption \citep{greene_thermal_2023,ih_constraining_2023}, but at the moment cannot rule out a hazy, thick, CO$_2$-rich atmosphere with a temperature inversion \citep{ducrot_combined_2024}. Secondary eclipse observations of TRAPPIST-1\,c with MIRI are also inconsistent with absorption from CO$_2$ \citep{zieba_no_2023,lincowski_potential_2023}. While some studies showed that a lack of atmosphere on the two innermost planets would not preclude atmospheres on the habitable-zone or outermost planets \citep{krissansen-totton_implications_2023,gialluca_implications_2024}, others suggest that it is unlikely that any significant atmosphere has survived on any of the TRAPPIST-1 planets \citep[e.g.,][]{van_looveren_airy_2024}. These two different conclusions are in part explained by differences in the rate at which the primordial magma ocean is solidified, which impacts the protected volatile reservoir.

In addition to eclipses, transits of all \hbox{TRAPPIST-1} planets were also observed \citep[][see also GO/GTO programs 1201, 1331, 1981, 2420, 2589, 6456]{lim_atmospheric_2023,radica_promise_2024,rathcke_stellar_2024}, but a challenge with transit spectroscopy of exoplanets orbiting active stars is the transit light source effect \citep[TLSE hereafter;][]{sing_hubble_2011,mccullough_water_2014,rackham_access_2017,rackham_transit_2018,rackham_sag21_2023}, i.e., contamination of the transit spectrum by stellar features caused by unocculted stellar spots and/or faculae. Such stellar contamination has been a source of uncertainty in the search for atmospheres on planets orbiting M dwarfs with JWST \citep[e.g.,][]{moran_high_2023,may_double_2023,lim_atmospheric_2023,cadieux_transmission_2024,radica_promise_2024,canas_GEMS_2026}, in some cases dominating the signal in transit spectra with signatures of hundreds of ppm. Although the TLSE is now commonly included in analyses of transit spectroscopy data of planets around potentially heterogeneous stars, some observations of such systems revealed no evidence for the TLSE \citep[e.g.,][also TRAPPIST-1\,g observations with NIRSpec, Benneke et al., under review]{lustig-yaeger_jwst_2023}. 

Observations of TRAPPIST-1 planets are also challenged by frequent, at times strong, stellar flares \citep[][also Benneke et al., in review]{luger_seven-planet_2017,vida_frequent_2017,ducrot_0845_2018,ducrot_trappist-1_2020,maas_lower-than-expected_2022,lim_atmospheric_2023,howard_characterizing_2023,radica_promise_2024,berardo_hubbles_2025,piaulet-ghorayeb_strict_2025}. 
When occurring near a transit, flares can lead to a time-dependent and chromatic dilution of the planetary transit, hence biasing the measured transit depth and consequently the transit spectrum and its interpretation. 

Despite its active and unpredictable nature, the host star of the TRAPPIST-1 system remains a prime target to search for atmospheres on Earth-sized exoplanet due to, among other things, increased planet-to-star radius ratios and shorter orbital periods. As one of the most favorable targets in the ``potentially habitable and Earth-sized'' regime \citep{gillon_trappist-1_2020}, TRAPPIST-1\,f was previously observed in transmission with the Hubble Space Telescope (HST), rejecting a cloud-free, hydrogen-dominated atmospheres at 4\,$\sigma$ \citep{de_wit_atmospheric_2018}. A joint analysis of broadband photometry from K2 and \textit{Spitzer} along with HST spectroscopy of TRAPPIST-1\,f \citep{ducrot_0845_2018} hints at some structure at the level of 200--300 ppm, which may be explained by the TLSE. A combined analysis of the spectra of planets b through g suggests that the planets are unlikely to host CH$_4$-rich atmospheres \citep{ducrot_trappist-1_2020}.

Here we present the first JWST observations of TRAPPIST-1\,f. The paper is structured as follows. We present the observations in Section~\ref{sec:observations} and describe the data reduction in Section~\ref{sec:reduction}. We discuss stellar flares and mitigation approaches in Section~\ref{sec:flares}. The transit light curve fitting is detailed in Section~\ref{sec:lcfit}, followed by an initial assessment of the transit spectra in Section~\ref{sec:tspec}. In Section~\ref{sec:retrieval} we present the planetary atmosphere and TLSE retrieval, and discuss the ensuing results in Section~\ref{sec:results}. We revisit the interior and habitability of TRAPPIST-1\,f in Section~\ref{sec:interior_habitability} and conclude in Section~\ref{sec:conclusion}.

\section{Observations} \label{sec:observations}

As part of the NIRISS Exploration of the Atmospheric diversity of Transiting exoplanets (NEAT; JWST GTO 1201, PI D. Lafreni\`{e}re), five transits of TRAPPIST-1\,f were observed on October 28$^{\rm th}$ 2022, June 15$^{\rm th}$ 2023, June 24$^{\rm th}$ 2023, July 3$^{\rm rd}$ 2023, and July 22$^{\rm nd}$ 2023 with NIRISS \citep{doyon_near_2023} in SOSS mode \citep{albert_near_2023}, covering 0.9--2.8\,$\mu$m at a resolving power $R\approx700$ (spectral order 1) and 0.6--1.4\,$\mu$m at $R\approx1400$ (spectral order 2). We note that the wavelength range of spectral order 2 includes the H$\alpha$ stellar line, allowing to monitor stellar activity such as flares. 
Each observation was performed in a single exposure of 121 integrations of 1.65\,min (18 groups per integration, duty cycle of 89.5\%), for a total of approximately 3.5\,h per visit, using the SUBSTRIP256 subarray and the NISRAPID readout pattern. In all visits, the median signal at the group level is 12.5\,kADU with a standard deviation of 3.4\,kADU and 99.99\% of the pixels below 55\,kADU, well below the 62--65\,kADU at which the onset of $\geq10\%$ non-linearity or saturation occurs \citep{albert_near_2023}. The time-medianed, per-integration signal-to-noise ratios (SNRs) in the extracted spectra are approximately 790, 580, and 360 in bands J, H and K, respectively, in all five visits. No apparent tilt events were seen in any of the visits, but asteroids crossed the detector during visits 1 and 3 (details in Appendix~\ref{app:reduction}).

\section{Data Reduction} \label{sec:reduction}

We reduced the data from the five visits with the \texttt{SOSSISSE} pipeline \citep[Appendices A of][]{lim_atmospheric_2023,cadieux_transmission_2024}. 
We also reduced the data with the \texttt{exoTEDRF} pipeline \citep[formerly \texttt{supreme-SPOON,}][]{feinstein_early_2023,coulombe_broadband_2023,radica_awesome_2023,radica_exotedrf_2024}. The differences between the transit spectra from \texttt{exoTEDRF} and \texttt{SOSSISSE} are, on average, smaller than the transit depth uncertainties (Figure~\ref{fig:tspec_pipelines}). The largest differences between reductions are in Visit~4, which is also the only visit where the two reductions lead to slightly different overall slopes in the spectra. Unless otherwise specified, we used the \texttt{SOSSISSE} pipeline for the rest of this work. Extracted spectra and other data products can be found on Zenodo via\dataset[DOI: 10.5281/zenodo.14501417]{https://doi.org/10.5281/zenodo.14501417}.

\section{Stellar flares} \label{sec:flares}

\subsection{Identification} \label{ssec:flares_id}

\begin{figure*}
    \centering
    \includegraphics[height=.37\textheight]{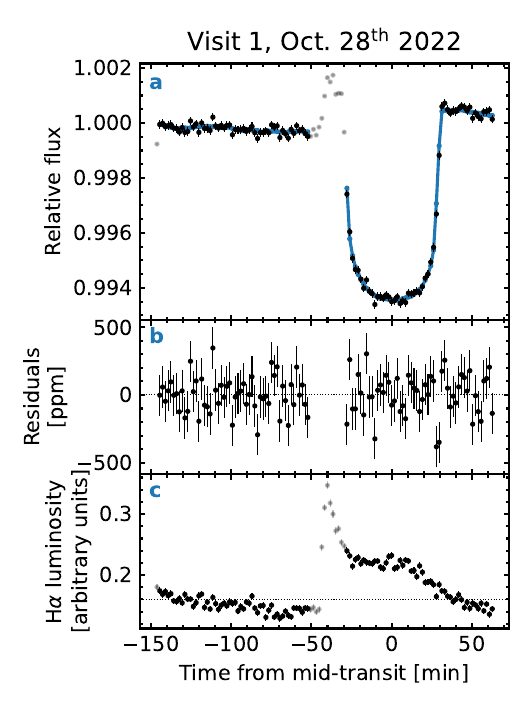}
    \includegraphics[height=.37\textheight]{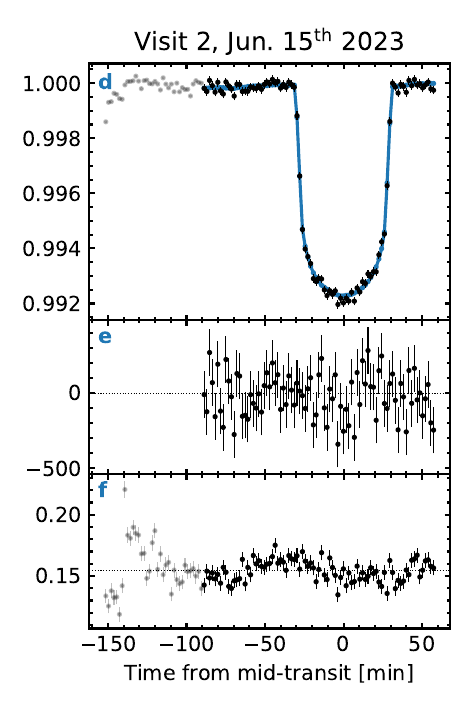}
    \includegraphics[height=.37\textheight]{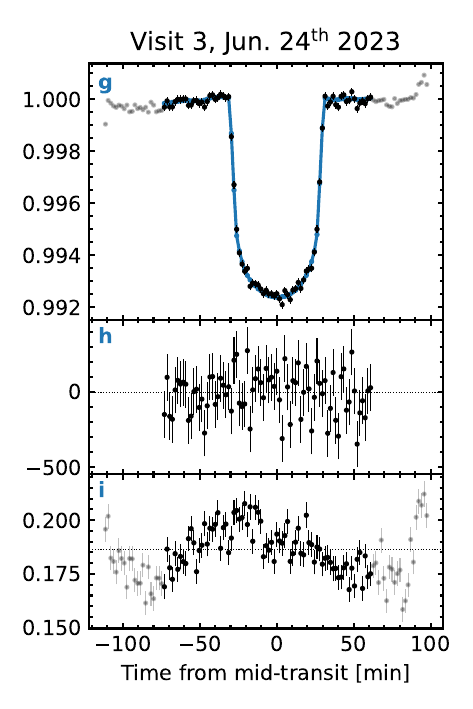}
    \includegraphics[height=.37\textheight]{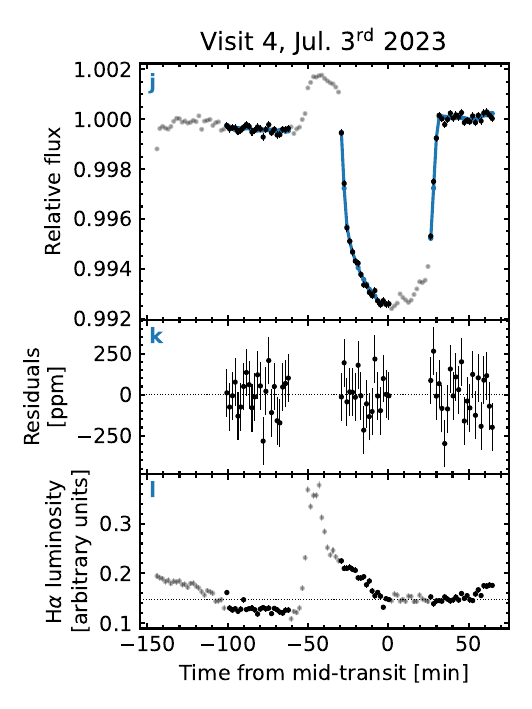}
    \includegraphics[height=.37\textheight]{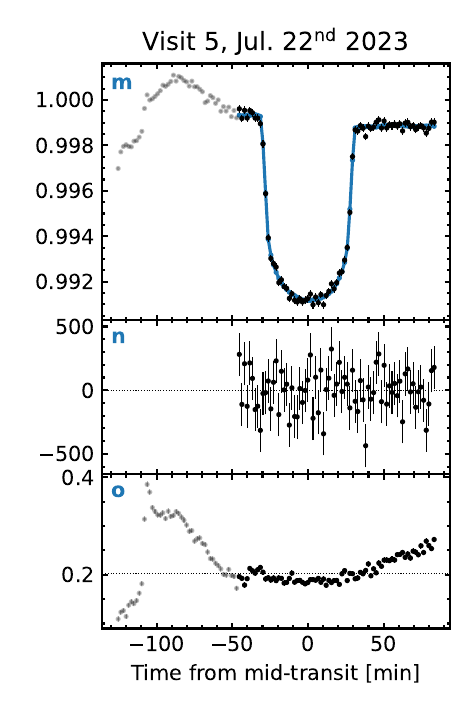}
    \caption{
    The top panel of each subplot shows the broadband light curve (black and gray points) and the best-fit transit and systematics model (blue curve) for the five transits of TRAPPIST-1\,f. Gray points are integrations that were masked in the light curve fit. The in-transit, masked integrations in Visit~4 correspond to a possible spot-crossing event. All other masked integrations are masked due to stellar activity. 
    The middle panels show the residuals from the best-fit model while the bottom panels show the integrated flux of the H$\alpha$ stellar line. Gray points correspond to the light curve points that were masked in the fit. All error bars are 1-$\sigma$ uncertainties.}
    \label{fig:lcfit}
\end{figure*}

At least one stellar flare occurred in each of the five visits. We identify an event as a flare when there is a rapid increase followed by a more gradual decrease in flux in the broadband light curve, and this trend must be aligned in time with a similar increase-decrease in the time series of the integrated flux of the H$\alpha$ stellar line (bottom panel of each visit in Figure~\ref{fig:lcfit}). Visit~1 has a relatively high-energy flare with a flux at its peak equivalent to about $30\%$ of the transit depth starting about 12\,min before the transit ingress, with the decreasing phase of the flare overlapping the transit. In Visit~2, a lower-energy flare ($\sim13\%$ of the transit depth) occurred about 1\,hour 48\,minutes prior to the ingress. In Visit~3, a similar low-energy flare ($\sim13\%$ of the transit depth) occurred about 54~minutes \textit{after} the transit egress. Similar to Visit~1, Visits~4 and 5 each have a high-energy flare ($\sim25\%$ of the transit depth) starting about 23\,min and 1~hour 17~minutes before the transit ingress, respectively. The flare certainly overlaps the transit in Visit~4 and likely does so in Visit~5 as well given the flare amplitude. 

The flare in Visit~1 was presented in detail in \cite{howard_characterizing_2023}. For an in-depth spectroscopic analysis of the other flares, we refer to \citet{howard_separating_2025,vasilyev_flares_2025,vasilyev_single_2026}. Here, we focus on how to empirically mitigate the dilution effect of these flares on the transit depth measurements.

\subsection{Mitigation} \label{ssec:flares_mitigation}

For all visits, we masked a 4-nanometer region in all spectra around each of the following spectral lines to remove any possible flare emission lines: Balmer $\alpha$ (H$\alpha$), the first six lines of the Paschen series, the first five lines of the Brackett series, and the helium triplet near 1.083\,$\mu$m. Only the largest flares showed emission in all or most of these lines -- smaller flares did not, but we masked all these lines in all flares out of caution. We then looked at how to further mitigate the effects of the flares in each visit.

The easiest flare to handle is the one in Visit~3 because it occurred after the transit and it should therefore not contribute any flux during the transit. However, the flare can bias the post-transit baseline and, as a consequence, the transit depth measurement. Luckily, the onset of a flare is easier to locate in time than the end of a flare. We thus only kept integrations up to 30~minutes after the egress. Keeping integrations up to about 50~minutes after the egress does not significantly affect the transit spectrum. The pre-transit masking is discussed in Section~\ref{sec:lcfit}.

The second easiest flare to treat is the one in Visit~2 because it emitted little energy and occurred long enough before the ingress that it likely contributes little to no flux during the transit. We fitted the light curves of Visit~2 by keeping either about 1\,hour~15\,minutes, 1\,hour, or 30\,min of the baseline immediately preceding the ingress, and there is no significant difference between the three resulting transit spectra. We opted for the 1-hour baseline to balance flare removal and accurate baseline measurement.

Although the flare in Visit~5 was highly energetic and occurred before the transit, we attempted to salvage this visit by removing integrations preceding the transit. In one case, we removed about 1\,hour~20\,minutes of baseline starting from the beginning of the visit, leaving about 14~minutes of pre-ingress baseline, and in the other case, we removed 2~hours of baseline, removing all pre-ingress baseline as well as the first 45\% of the transit. The resulting transit spectra do not agree as well as the different scenarios of Visits~2 and 3, but it is unclear whether the differences are explained by a worse removal of the flare in the first case, or by a more difficult determination of the transit depth in the second case, or both. We note that during and after the transit, the integrated flux in the H$\alpha$ stellar line is smoothly increasing (Figure~\ref{fig:lcfit}, panel o). Assuming that the stellar flux at other wavelengths behaves similarly to the H$\alpha$ line, we opted for the 1-hour-20-minute baseline cut, assuming that the additional light curve treatments discussed in Section~\ref{sec:lcfit} will catch any remaining stellar variability.

Visits~1 and 4 are the two visits most affected by stellar flares because in both visits the flares occurred shortly before ingress. If we were to simply mask the flare, it is unclear which integrations we should mask \textit{during} the transit. In fact, the flare likely contributes to the flux significantly even at the egress and later: the post-transit baseline is not aligned with the pre-transit, pre-flare baseline. We tested an iterative flare correction approach in which we alternate between subtracting a best-fitting, temporally cooling, blackbody model out of the spectral time series, and fitting those ``flare-corrected'' light curves with a transit model (details in Appendix~\ref{app:flarecorr}). The resulting transit spectra exhibited features similar to the spectra obtained \textit{without} any flare correction, and those features are unlikely to be explained by the TLSE or by a planetary atmosphere, so we did not use this correction approach in the end. 
We instead masked the integrations that are clearly during the flare and before the ingress, and combined this masking with the additional light curve treatments discussed in Section~\ref{sec:lcfit}. This approach at least provides more reasonable transit spectra, and the details of how these transit spectra are included in the retrieval are given in Section~\ref{ssec:retrieval_multivisit}.

We also simulated a spectral time series with a transit and a flare injected near the ingress (Appendix~\ref{app:flaresim}), similar to Visits~1 and 4, to see the effect of the flare on the transit spectrum, but we find that the simulations cannot reproduce the observed transit spectra, likely due to the simplicity of the flare (and/or noise) model. However, these simulations showed that flares can mimic the signal of unocculted stellar faculae, meaning that faculae-like signatures in a transit spectrum could in fact be due to uncorrected stellar flares or, more generally, uncorrected stellar variability. Using a TLSE model to correct for such signatures may thus be conceptually flawed, because flares tend to evolve on timescales of the order of minutes in terms of amplitudes and temperatures, while TLSE models assume the heterogeneities are static throughout the observation, which typically lasts several hours. 

\section{Light curve fitting} \label{sec:lcfit}

With the spectral time series of each visit, we first generated a broadband light curve by summing all the fluxes of spectral order 1 (0.85 -- 2.85\,$\mu$m) along the wavelength direction (top panel of each visit in Figure~\ref{fig:lcfit}). We also produced a set of spectroscopic light curves by binning the fluxes into 49 bins that are equally spaced in wavelength (Figure~\ref{fig:slcs_residuals}). For spectral order 2, we binned all the fluxes into a single wavelength bin due to the low stellar flux and lower instrumental throughput at these wavelengths. 
We performed a 10-iteration 5-sigma clip on the broadband and spectroscopic light curves. We then fitted the light curves with \texttt{ExoTEP} \citep{benneke_spitzer_2017,benneke_water_2019,benneke_sub-neptune_2019}, which includes the \texttt{batman} transit model \citep{kreidberg_batman_2015} along with custom systematics models described later in this section. The light curves are fitted via MCMC, implemented with \texttt{emcee} \citep{foreman-mackey_emcee_2013}, with 10\,000 steps (6\,000 burn-in + 4\,000 production) and the number of walkers equal to four times the number of free parameters. 

The broadband light curves were first fitted to find the optimal orbital parameters, which should be common across all five visits. 
The spectroscopic light curves were then fitted with the orbital parameters \textit{fixed} to the optimal values found with the broadband light curves, but with the limb darkening coefficients free. This first iteration of spectroscopic light curve fits yielded optimal limb darkening coefficients for each visit, which we error-weighted averaged with the optimal coefficients from two visits of TRAPPIST-1\,b and two of TRAPPIST-1\,c with NIRISS/SOSS \citep[][see Figure~\ref{fig:ldcs_spread} for the error-weighted averaged limb darkening coefficients]{lim_atmospheric_2023,radica_promise_2024} to find the optimal spectroscopic limb darkening coefficients based on as many JWST/SOSS/TRAPPIST-1 visits as possible. The limb darkening coefficients were then fixed to these error-weighted average values in the final fit of the spectroscopic light curves. 
Fixing the LDCs may artificially reduce the transit depth uncertainty (Figure~\ref{fig:tspec_upper-lower-ldcs}). We discuss how fixing the LDCs may affect our conclusions in Section~\ref{app:lcfit_paramfix} and \ref{ssec:caveats}). 
Additional details on the iterative light curve fit, including fixing of the orbital and limb darkening parameters, are provided in Appendix~\ref{app:lcfit_paramfix}, with priors listed in Table~\ref{tab:priors_posteriors_lcfit}.

To account for possible stellar variability \citep[see H$\alpha$ variability in bottom panel of each visit in Figure~\ref{fig:lcfit}; see also][]{berardo_hubbles_2025}, we used three systematics models, all fitted simultaneously with the transit model for the broadband light curve, each spectroscopic light curve, and the order 2 light curve: a linear function, a linear detrending against the H$\alpha$ integrated flux time series, and a simple-harmonic-oscillator-kernel Gaussian process (Appendix~\ref{app:lcfit_sysmodels}, Appendix~B.3 of \citet{lim_atmospheric_2023}, and, e.g., \citet{radica_muted_2024,coulombe_highly_2025} for the use of this specific Gaussian process kernel). 

For Visit~3, in addition to removing the flare integrations (Section~\ref{ssec:flares_mitigation}), we also removed the integrations from the beginning of the visit to 45~minutes prior to ingress due to variability in the H$\alpha$ stellar line (Figure~\ref{fig:lcfit}, panel i). It is unclear whether this H$\alpha$ variability is caused by a flare, but we removed this part of the observation to avoid biasing the baseline. We also tried removing integrations 1~hour prior to ingress instead of 45~minutes and there was no significant impact on the transit spectrum. For the same reason, in Visit~4, we removed the first 44~minutes of the observation (Figure~\ref{fig:lcfit}, panels j and l). We also removed 14 in-transit integrations (approximately 24~minutes) due to a possible spot-crossing event.

To summarize, for all visits, the light curve fit is performed with the LDCs fixed to common values; the systematics model consists of a linear function, a GP, and a detrending against H$\alpha$. Each visit additionally has several integrations and/or pixels masked depending on the events that occurred in those visits, be it a flare, (non-flare) stellar variability, a possible spot-crossing event, and/or an asteroid crossing the detector.

\begin{figure*}
    \centering
    \includegraphics[trim={0 1.5em 0 0}, width=.49\textwidth]{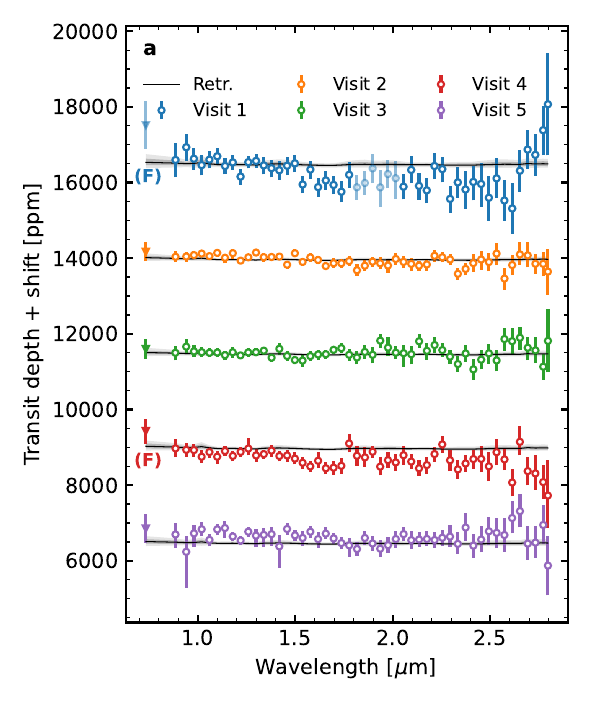}
    \includegraphics[trim={0 1.5em 0 0}, width=.49\textwidth]{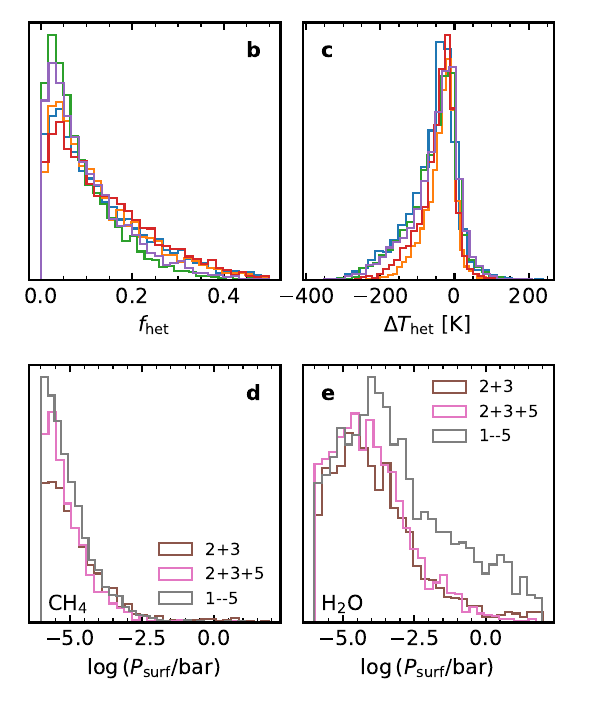}
    \caption{
    (a) Transit spectrum of each of the five TRAPPIST-1\,f visits, with vertical offset. Spectra are derived from systematics-corrected light curves. Points from spectral order 1 are represented as empty circles and spectral order 2 is represented as a filled triangle. Paler points in Visit~1 were ignored due to an asteroid crossing (details in Appendix~\ref{app:reduction_asteroid}). The ``(F)'' on the left indicates visits with a flare near ingress. Black curves and grey shaded regions show the retrieved (median, $\pm1\sigma$, and $\pm2\sigma$) N$_2$-dominated atmosphere and TLSE models from the simultaneous fit of Visits~1--5, i.e., five visits fitted with different TLSE parameters but a common planetary radius and quiet photosphere temperature. 
    (b) \& (c) Posterior distributions of the stellar heterogeneity covering fraction and temperature difference with respect to the quiet photosphere, assuming a haze-free, N$_2$-dominated atmosphere. Each histogram corresponds to a visit, with the same color code as in panel (a). These distributions were obtained from the same simultaneous fit of Visits~1--5 as for the retrieved models in panel a.
    (d) \& (e) Posterior distributions of the (log) surface pressure of pure CH$_4$ (d) and pure H$_2$O (e) \textit{haze-free} atmospheres when combining Visits~2+3 (brown), 2+3+5 (pink), or 1--5 (gray). 
    Transit spectra and other data products can be found on Zenodo via\dataset[DOI: 10.5281/zenodo.14501417]{https://doi.org/10.5281/zenodo.14501417}.
    }
    \label{fig:tspec_retrieval-posteriors}
\end{figure*}

\begin{figure*}
    \centering
    \includegraphics[width=\textwidth]{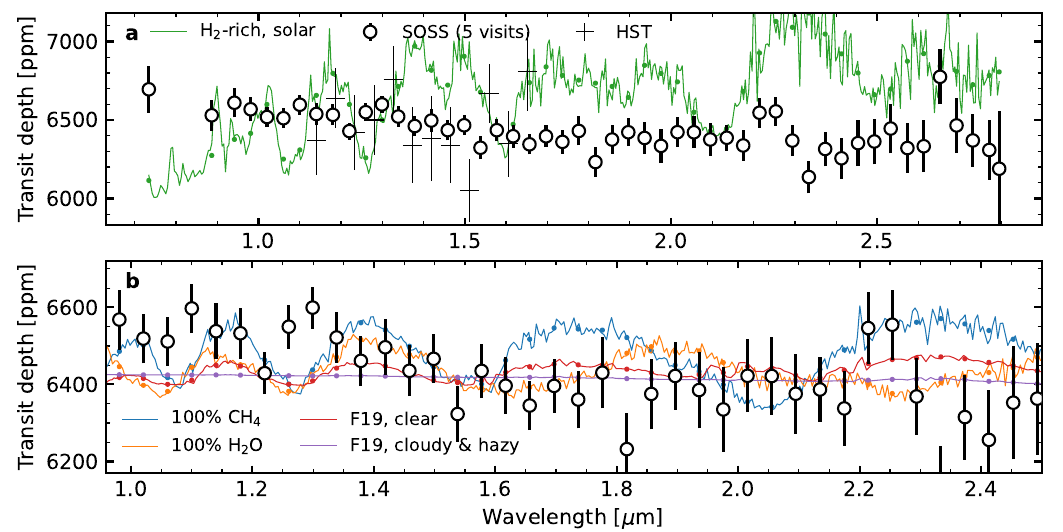}
    \caption{
    (a) Error-weighted average of the five TLSE-corrected SOSS transit spectra of TRAPPIST-1\,f (black points). The green curve is a H$_2$/He-dominated, 1-bar atmospheric model at solar metallicity. Green points are at the resolution of the data. Crosses are HST data \citep{de_wit_atmospheric_2018,zhang_near-infrared_2018}. 
    (b) Same SOSS spectrum as in panel (a), zoomed in, compared to pure, 1-bar atmospheric models (blue and orange), and to the Archean-Earth models from \citet[][red and purple]{fauchez_impact_2019}. 
    This transit spectrum and other data products can be found on Zenodo via\dataset[DOI: 10.5281/zenodo.14501417]{https://doi.org/10.5281/zenodo.14501417}.
    }
    \label{fig:tspec-stacked_models-overplotted}
\end{figure*}

\section{Initial Assessment of the Transit Spectra} \label{sec:tspec}

In this section, we describe features seen in the transit spectra of the five visits (Figure~\ref{fig:tspec_retrieval-posteriors}a) and we provide general statistics to assess the quality of the spectra. The median error bars on the transit depths are 346, 146, 186, 232, and 226\,ppm for Visits~1 to 5, respectively. By fitting a linear function to each transit spectrum, we obtained slopes of $-365^{+95}_{-100}$, $-185^{+43}_{-41}$, $+15^{+50}_{-50}$, $-208^{+61}_{-60}$, and $-98^{+69}_{-69}\,$ppm/$\mu$m for Visits~1 to 5, respectively. The treatment and interpretation of these slopes is detailed in Sections~\ref{ssec:retrieval_atm} and \ref{ssec:results_atm_other}.

Combining an increasing number of TLSE-corrected (Section~\ref{ssec:results_TLSE}) transit spectra with an error-weighted average reduces the dispersion in the combined transit spectrum (i.e., the standard deviation of the transit depths) by a factor of $\sqrt{N}$, where $N$ is the number of combined spectra, indicating that the spectra are dominated by random noise, not by structures correlated from one visit to another. The dispersion decreasing with $\sqrt{N}$ may be expected for planets orbiting a less active star, but confirming this trend with \hbox{TRAPPIST-1} is important considering the amount of stellar activity observed in these five visits and in other \hbox{TRAPPIST-1} observations. 

The transit spectrum of Visit~1 has the largest uncertainties and the steepest slope, which we attribute to the presence of the flare shortly before ingress. The \hbox{2000-ppm} increase in transit depth at long wavelengths is most likely not planetary in origin and due to degeneracies between $(R_p/R_\star)^2$ and the GP parameters and/or the H$\alpha$ detrending parameter, i.e., stellar variability that is not properly accounted for. We tried, for instance, masking fewer integrations around the flare, not masking the flare, not detrending against the H$\alpha$ flux, but we could not remove this increase without leaving a \hbox{1000--1500\,ppm} increase bluewards of $\sim$$1.7\,\mu$m. The increase at long wavelengths is also somewhat present in the transit spectrum resulting from the \texttt{exoTEDRF} reduction (Figure~\ref{fig:tspec_pipelines}). Since this increase affects only five bins which already have large error bars due to the low NIRISS/SOSS throughput, we ran the retrievals described in Section~\ref{sec:retrieval} both with and without those five affected bins and found that they have little to no effect on the conclusions.

The transit spectra of Visits~2 to 5, although relatively flat compared to that of Visit~1, each have unique features (e.g., near 2.3\,$\mu$m in Visit~2, 1.5--2.5\,$\mu$m in Visit 3, 1.8--2.3\,$\mu$m in Visit~4, 1.65--2\,$\mu$m in Visit~5). Given that these features have amplitudes of 200--500\,ppm, they are also unlikely to be planetary in origin and are probably due instead to stellar variability, instrument systematics, and/or data reduction methodologies. We note that these characteristics are also generally present in the spectra from the \texttt{exoTEDRF} pipeline (Figure~\ref{fig:tspec_pipelines}). 

\section{Retrievals} \label{sec:retrieval}

The transit spectra are fitted with the \texttt{POSEIDON} atmospheric retrieval code \citep{macdonald_hd_2017,macdonald_poseidon_2023}, which can fit a planetary atmosphere and/or the TLSE to the data \citep{fournier-tondreau_near-infrared_2024}. The TLSE component of the model can include either one heterogeneity, which can be colder or hotter than the quiet photosphere, or two heterogeneities, one colder and one hotter than the quiet photosphere. \texttt{POSEIDON} can also fit multiple visits simultaneously, forcing common planetary atmosphere parameters across all visits while allowing for different TLSE parameters for different visits. All fits were performed by nested sampling implemented with \texttt{PyMultiNest} \citep{feroz_multinest_2009,Buchner2014} with 400 live points and a (log) Bayesian evidence tolerance of 0.5.

\subsection{Multi-Visit Fit Considerations} \label{ssec:retrieval_multivisit}

Since the five visits span a range of contamination from stellar flares, we ran retrievals with different combinations of transit spectra. We first ran retrievals with each transit spectrum individually to see what information is provided by each visit. We then ran the same retrievals fitting Visits~2 and 3 simultaneously (2+3 hereafter), the two visits least affected by flares, then with Visits~2, 3, and 5 (2+3+5), and finally with all five visits (1--5). This cumulative approach allows us to assess how including more transits may help constrain the fitted parameters, while remaining cautious when interpreting any results involving flare-contaminated transits. When fitting two or more transit spectra simultaneously, we first shifted all the spectra to the same median to account for offsets between visits. 

For each combination of fitted visit(s), we first fitted the simplest model, a constant, with the planetary radius as the only free parameter. We then explored more complex models described in the following two subsections.

\subsection{TLSE Retrievals} \label{ssec:retrieval_TLSE}

Given the detection of the TLSE in other transit spectra of TRAPPIST-1 planets \citep[e.g.,][]{lim_atmospheric_2023,radica_promise_2024}, we fitted the spectrum (or spectra) of each combination of visits with a single-heterogeneity TLSE model. In addition to the planetary radius, we fitted for the temperature of the quiet photosphere with a Gaussian prior, and fitted for the covering fraction and the temperature of the heterogeneity with uniform priors (Table~\ref{tab:retrieval_priors}). \texttt{POSEIDON} allows including two heterogeneities instead of only one, and also fitting for the surface gravities of the quiet photosphere and those of the heterogeneities, but we tried fitting for such additional parameters and the data did not show a statistical preference for these more complex models. 

\subsection{Planetary Atmosphere Retrievals} \label{ssec:retrieval_atm}

We fitted the spectrum/spectra of each combination of visits with an isothermal, H$_2$- and He-dominated or N$_2$-dominated atmospheric model with CH$_4$ \citep{yurchenko_exomol_2024}, H$_2$O \citep{polyansky_exomol_2018}, CO$_2$ \citep{yurchenko_exomol_2020}, CO \citep{li_rovibrational_2015}, NH$_3$ \citep{coles_exomol_2019}, and PH$_3$ \citep{sousa-silva_exomol_2015} as trace species, as well as N$_2$ in the H$_2$/He-dominated scenario and H$_2$ in the N$_2$-dominated scenario. The free atmospheric parameters are the volume mixing ratios of the trace species, the temperature, the reference radius corresponding to a pressure of 1\,bar, and the surface pressure (Table~\ref{tab:retrieval_priors}). 

Additionally, we fitted the spectrum/spectra of each combination of visits with an isothermal pure (single-species) atmospheric model of CH$_4$, H$_2$O, CO$_2$, CO, NH$_3$ or SO$_2$ to see if we could put any upper limits on the surface pressures. 
We also fitted the spectra with an isothermal, N$_2$-dominated atmospheric model containing only one of either CH$_4$, H$_2$O, CO, or NH$_3$ as a trace species \citep{radica_promise_2024,piaulet-ghorayeb_strict_2025}.

We ran all retrievals with and without a single-heterogeneity TLSE model. Unless otherwise stated, all results presented hereafter include a TLSE model. 

Given the generally negative slopes in the transit spectra (Section~\ref{sec:tspec}), we additionally tried to include hazes in all retrievals, thus adding free parameters $a$ and $\gamma$, the ``Rayleigh-enhancement factor'' and ``scattering slope'' from \citet{macdonald_hd_2017}, based on \citet{lecavelier_des_etangs_rayleigh_2008}. 

\section{Results and Discussion} \label{sec:results}

\subsection{No Evidence of the TLSE} \label{ssec:results_TLSE}

Across all combinations of visits and all models explored, we found no evidence of the TLSE, either from model comparison via the Bayesian evidences \citep[all confidence levels stated in numbers of $\sigma$ are computed with Equation~(27) from][]{trotta_bayes_2008} or from the inferred heterogeneity covering fractions and temperatures.

Regardless of whether we assume TRAPPIST-1\,f has an atmosphere -- and irrespective of the assumed dominant species -- comparing the Bayesian evidence of the model with and without TLSE shows no preference for including TLSE, regardless of the combination of visit(s) considered (see rejection confidence levels in Table~\ref{tab:model_comparison}). The models with TLSE are rejected at the 2--3$\sigma$ level for individual visits, and the confidence level increases as more visits are combined, up to around $6\sigma$ when fitting Visits~1--5. If we assume that all visits share the same TLSE parameters, the TLSE remains undetected at the $3\sigma$ confidence level. The posterior distributions of the heterogeneity covering fractions and of the temperature differences between the heterogeneity and the quiet photosphere show a preference for small (near-zero) values. For example, panels b and c of Figure~\ref{fig:tspec_retrieval-posteriors} show the posterior distributions of the covering fractions and heterogeneity temperature differences from the fit of Visits~1--5, assuming a haze-free, N$_2$-dominated atmosphere with multiple trace species. This is in stark contrast with previous JWST observations of inner TRAPPIST-1 planets, which found covering fractions of 20--30\% and temperature differences of 150--200\,K for TRAPPIST-1\,b and c \citep{lim_atmospheric_2023,radica_promise_2024}. Our conclusions for TRAPPIST-1\,f remain unchanged if we remove spectral order 2, or if we remove the five reddest bins in the transit spectrum of Visit~1, or if we use instead the \texttt{exoTEDRF} reduction pipeline, and whether hazes are included or not. 

Nevertheless, we still include a TLSE model in all retrievals to ensure that all results are marginalized over any possible TLSE signature (see also possible caveats in the analysis; Section~\ref{ssec:caveats}). We also use the retrieved TLSE spectra to correct the observed spectra from any potential TLSE whenever transit spectra are combined, e.g., in panels a and b of Figure~\ref{fig:tspec-stacked_models-overplotted}. This TLSE correction is done by dividing out the retrieved median TLSE spectrum from each transit spectrum and propagating the 1-$\sigma$ uncertainty on the retrieved TLSE spectrum (which accounts for uncertainties on the TLSE stellar parameters) onto the transit depths.


If we exclude Visit~1 from the following discussion, considering it took place about a year before the other four visits, and assuming that there was truly no TLSE during Visits~2 to 5, we could conclude that there was no TLSE in four visits spread over a little over a month (mid-June to end of July 2023), which roughly corresponds to ten stellar rotations \citep[adopting a stellar rotation period of 3.3\,days,][]{luger_seven-planet_2017,vida_frequent_2017,berardo_hubbles_2025}. The apparent absence of TLSE raises some questions: Did the star happen to have fewer heterogeneities on those four visits compared to 
other visits targeting other planets in the system \citep[e.g.,][]{lim_atmospheric_2023,radica_promise_2024,piaulet-ghorayeb_strict_2025}? 
How is the presence of stellar surface heterogeneities correlated with the presence of flares? 

An absence of TLSE can also be explained by the transit chord having similar properties as the average, out-of-transit, visible stellar hemisphere, which would not require the star to have a homogeneous surface \citep[][their Figure~4]{rathcke_stellar_2024}. 
The lack of TLSE in the TRAPPIST-1\,f visits is consistent with NIRSpec observations of TRAPPIST-1\,g (17 July and 12 December 2022, Benneke et al., in review), and these two planets have similar impact parameters \citep[$b\approx0.312$ and $0.379$ for planets f and g, respectively][]{agol_refining_2021}. How does the TLSE vary with the latitude of the transit chord? Given the detection of the TLSE in closer-in planets b, c, and d \citep{lim_atmospheric_2023,radica_promise_2024,piaulet-ghorayeb_strict_2025}, could the transit chords of TRAPPIST-1\,f and g have properties more similar to the average, out-of-transit, visible stellar disk than the chords of the closer-in planets \citep[$b\approx0.095$, $0.109$, and $0.063$ for planets b, c, and d, respectively,][]{agol_refining_2021}? For example, if the stellar surface has very few spots at its equator and progressively more spots towards its poles, then transit chords at the stellar equator (and poles) would differ more from the average, out-of-transit, visible stellar disk than the transit chords at intermediate impact parameters. In that case, the transit chords of planets f and g should include more stellar heterogeneities than those of the closer-in planets and one could then ask why we see so few spot-crossing events. If the heterogeneities are very small (but numerous), the signal-to-noise ratio in the light curves might be insufficient to detect the crossing of such spots/faculae. If higher-impact-parameter planets are indeed less affected by the TLSE, we should see a difference in TLSE signatures in the transit spectra of back-to-back transits of low- and high-impact-parameter planets, but this would also complicate the strategy of using the TLSE signature from transits of low-impact-parameter planets to correct the TLSE at higher impact parameters \citep{trappist-1_jwst_community_initiative_roadmap_2024,allen_using_2024,rathcke_stellar_2024}. 
Addressing all these questions is outside the scope of this work, but given the importance that was given to the TLSE in previous works on small planets orbiting M dwarfs \citep[e.g.,][]{lim_atmospheric_2023,moran_high_2023,may_double_2023,cadieux_transmission_2024,radica_promise_2024}, we list these questions here to add some nuances to upcoming discussions and works on such targets.

\subsection{No Thick H$_2$/He-Dominated Atmosphere} \label{ssec:results_atm_H2He}

The retrievals with a H$_2$/He-dominated atmosphere + TLSE constrain the surface pressure to $P_{\rm surf}\lesssim3\,$mbar at 95\% confidence based on Visits~1--5, in good agreement with the conclusion from \citet{de_wit_atmospheric_2018}. This upper limit improves as more visits are combined (Table~\ref{tab:model_comparison}), is broadly consistent across the two reduction pipelines, and remains unchanged if we remove the five reddest bins in Visit~1. Including hazes increases the upper limit to 13\,bar (Visits~1--5), but hazes are only detected (\hbox{3--4\,$\sigma$}, based on Bayesian evidences) in Visits~1, 2, and 4, and different ``slope'' parameters ($\gamma$) are inferred for each of these three visits. Fitting only Visits~3 and 5 gives an upper limit of 16\,mbar. Panel a of Figure~\ref{fig:tspec-stacked_models-overplotted} compares the combined (five-visit, TLSE-corrected) transit spectrum of TRAPPIST-1\,f to a 1-bar, H$_2$- and He-dominated atmosphere at solar metallicity, rejected at 85\,$\sigma$.

\subsection{Constraints on High-Mean-Molecular-Mass Atmospheres} \label{ssec:results_atm_other}

We could not constrain the surface pressure of the N$_2$-dominated atmosphere with multiple trace species with either of the two reduction pipelines (Table~\ref{tab:model_comparison}). Although the two reductions agree in their absence of constraints in this specific case, we note that for all other high-mean-molecular-mass atmosphere scenarios presented in this subsection, the upper limit on the surface pressure, if there is one, depends on the chosen reduction pipeline, even if the transit spectra appear to be in good agreement (Figure~\ref{fig:tspec_pipelines}). This may be due to retrievals being highly sensitive to small variations in transit spectra \citep[e.g.,][]{constantinou_early_2023}. The upper limits derived from the \texttt{SOSSISSE} pipeline are generally more restrictive (lower surface pressures are preferred) than those from \texttt{exoTEDRF}. We present upper limits here, in panels d and e of Figure~\ref{fig:tspec_retrieval-posteriors}, and in Table~\ref{tab:model_comparison}, but we emphasize that these results should be taken in the context that they are sensitive to the reduction procedure.

For the pure (single-species) atmospheres, combining Visits~1--5 constrains the surface pressure to values lower than 0.52\,mbar, 4.9\,bar, and 1.7\,bar for CH$_4$, H$_2$O, and NH$_3$, respectively (Table~\ref{tab:model_comparison}; 95\% confidence level), with no relevant constraints on pure CO$_2$, CO, or SO$_2$ atmospheres. However, when fitting the transit spectra from \texttt{exoTEDRF} instead, we only obtain an upper limit of 8.8\,bar for H$_2$O, and no constraints on all other tested species. One should note though that if we only fit Visits~2 and 3 (the visits least affected by flares) of \texttt{exoTEDRF}, we then get an upper limit of 8.2\,bar on the pure CH$_4$ atmosphere, which is still much less constraining than with \texttt{SOSSISSE}, but might hint at problems specific to visits affected by flares. No constraints were inferred from the retrievals with N$_2$ + one trace species.

Including hazes invalidates most upper limits (Table~\ref{tab:model_comparison}). However, looking at individual visits, only Visit~2 has a strong detection (3--3.5\,$\sigma$) of hazes. The haze detection significance decreases when Visits~3 and 5 are combined, but increases when Visits~1 and 4 are combined. The slopes in the transit spectra are an order of magnitude larger than expected from realistic models \citep[e.g.,][]{fauchez_impact_2019}. We thus suspect that the slopes are largely, if not entirely, due to uncorrected stellar activity and not hazes (see also Appendix~\ref{app:flaresim}). Still, considering the haze parameters as nuisance parameters, we stress that the upper limits can vary significantly whether we account for residual slopes in the transit spectra or not.

The overall lower sensitivity to higher-mean-molecular-mass atmospheres can be seen in panel b of Figure~\ref{fig:tspec-stacked_models-overplotted}, where we compare the combined transit spectrum of TRAPPIST-1\,f to a selection of models. The blue (orange) curve is a pure CH$_4$ (H$_2$O), 1-bar atmospheric model, which is rejected at $6.8\sigma$ ($3.7\sigma$). The red (purple) curve is the clear-sky (cloudy-and-hazy), Archean-Earth model from \citet[][case B; see also \citet{charnay_exploring_2013}]{fauchez_impact_2019}, which is rejected at $4.3\sigma$ ($2.9\sigma$)\footnote{The combined transit spectrum confidently rejects these N$_2$-dominated atmosphere, but each transit spectrum was corrected for the TLSE prior to the combination, thus removing a maximal amount of features in the spectrum, whereas in the retrievals involving N$_2$-dominated atmospheres, we \textit{simultaneously} fit the atmosphere and the TLSE.}. The uncertainties on the combined transit spectrum are comparable to or larger than most expected spectral features.

\subsection{Possible caveats in the analysis}\label{ssec:caveats}

As shown in Figure~\ref{fig:tspec_upper-lower-ldcs}, fixing the limb darkening coefficients instead of fitting them may lead to underestimated transit depth uncertainties. This is especially true for spectral order 2. A larger uncertainty in order 2 may lead to a less confident non-detection of the TLSE because one of the main signatures of the TLSE is a slope in the blue end of the SOSS domain. However, we remind the reader that we ran retrievals with and without including spectral order 2 and consistently found no evidence for the TLSE (both in terms of Bayesian evidences and in terms of stellar heterogeneity parameters), suggesting that order 2 contributes little to our conclusions on the TLSE.

Fitting the light curves simultaneously with the transit model and the systematics models (linear function, GP, and H$\alpha$ detrending) may have biased the transit spectra. This is especially true for the GP and the H$\alpha$ detrending because they can model non-linear temporal variability, meaning that the parameters of these systematics models are more likely to be correlated with the planet-to-star radius ratio. The resulting transit spectra may in turn have affected our conclusions on the TLSE, i.e., the GP and/or the H$\alpha$ may have modeled out the TLSE signatures from the transit spectra.

We note however that a GP has been used with other JWST/SOSS/TRAPPIST-1 observations \citep{lim_atmospheric_2023,radica_promise_2024} and the TLSE \textit{is} detected in those cases. As for the H$\alpha$ detrending, Figure~\ref{fig:tspec_with-out-Ha} compares the transit spectra of Visits~1 and 2, each with and without detrending against H$\alpha$. Visit~1 has a flare shortly before the transit whereas Visit~2 does not. H$\alpha$ detrending has little to no effect on Visit~2 but in Visit~1, \textit{not} detrending against H$\alpha$ introduces a slope at wavelengths bluewards of $\sim$1.7\,$\mu$m and it removes the increase at redder wavelengths. The difference between the two visits suggests that the H$\alpha$ detrending is removing signals created by the flare, as it is meant to do, confirming the relevance of this systematics model for flare-contaminated visits. However, it remains possible that the H$\alpha$ detrending removed the TLSE signal partially or completely as flares produce signals similar to that of an unocculted facula (panel a of Figure~\ref{fig:flaresim_tspec}). We note though that the TLSE is still rejected at 4.1\,$\sigma$ when considering only Visits~2+3, the visits least affected by flares, and at 4.8\,$\sigma$ with Visits~2+3+5.

\section{Interior Modeling and Water Inventory} \label{sec:interior_habitability}


\begin{figure*}
    \centering
    \includegraphics[width=\linewidth]{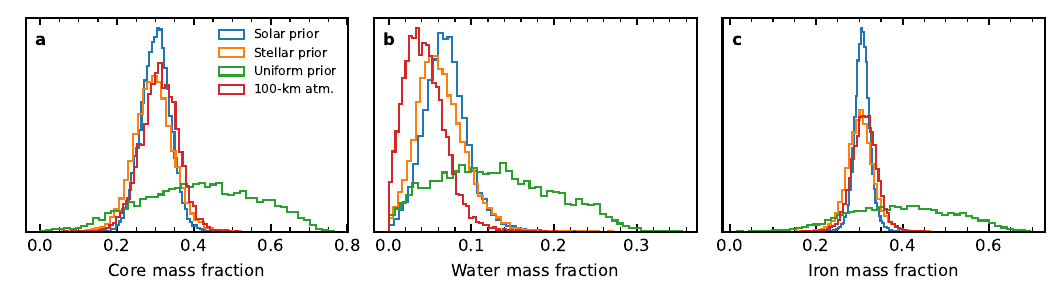}
    \caption{
    Posterior distributions of (a) the core mass fraction, (b) the water mass fraction, and (c) the iron mass fraction of TRAPPIST-1\,f, assuming either stellar priors (blue), solar priors (orange), or uniform priors (green) on the Fe/Si and Fe/Mg mass ratios. Red curves correspond to a model that includes a 100-km-high atmosphere. 
    }
    \label{fig:cmf_wmf}
\end{figure*}

M dwarfs remain in the pre-main-sequence phase longer than earlier-type stars, a phase during which the star is more luminous and active, thus subjecting the planets to extreme levels of radiation for longer periods of time \citep[e.g.,][and references therein]{luger_extreme_2015,turbet_review_2020}. For TRAPPIST-1, current XUV observations suggest that the planets were subject to persistent and strong XUV radiation since their formation \citep{wheatley_strong_2017,fleming_xuv_2020}, potentially leading to loss of volatile content. Models showed that TRAPPIST-1\,f may have lost $\sim4$ to 20 terrestrial oceans' (TO) worth of water, depending on assumptions and models \citep{bolmont_water_2017,bourrier_temporal_2017,gialluca_implications_2024}. 

Previous observations found the mass and radius of TRAPPIST-1\,f to be consistent with a range of interior compositions: from a water-free rocky composition with a core mass fraction (CMF) of $\sim18\%$, lower than Earth, to an Earth-like CMF \citep[$\sim32\%$,][]{wang_elemental_2018} with a $\sim5\%$ water mass fraction \citep[WMF; Table~9 of][]{agol_refining_2021}. 
Following the same methodology as in \citet{cadieux_toi-1452_2022,cadieux_new_2024} and using the new planetary radius inferred from the light curve fits (Table~\ref{tab:priors_posteriors_lcfit}), we constrain the CMF and WMF of TRAPPIST-1\,f. We use the planetary interior model from \citet{valencia_radius_2007} and \citet{plotnykov_chemical_2020}, which
assumes no water in the mantle or core, but does have variable amounts of light elements in the core (in the form of Si) and variable degrees of mantle differentiation (amounts of Fe in mantle). Assuming the refractory abundance ratios of the planet are either solar (i.e., solar priors are applied: ${\rm Fe/Mg}=1.87^{+0.23}_{-0.20}$ and ${\rm Fe/Si}=1.77^{+0.22}_{-0.19}$ by mass), stellar (${\rm Fe/Mg}=1.87^{+0.48}_{-0.37}$ and ${\rm Fe/Si}=1.77^{+0.45}_{-0.36}$), or applying uniform priors, we obtain CMF and WMF of $30\pm4\%$ and $7^{+3}_{-2}\%$, $30\pm5\%$ and $6\pm3\%$, and $38^{+18}_{-17}\%$ and $11^{+9}_{-7}\%$, respectively (Figure~\ref{fig:cmf_wmf}), broadly consistent with \citet{acuna_characterisation_2021} and the \citet{agol_refining_2021} scenario for an Earth-like CMF. Adding a 100-km-high atmosphere mainly impacts the WMF, weakening the constraint to an upper limit of $\mathrm{WMF}<8.7\%$ at 95\% confidence 
(red histograms in Figure~\ref{fig:cmf_wmf}). We note that the uniform-prior case leads to Fe/Mg and Fe/Si about 1.4 times larger than solar values. Forcing the WMF to zero leads to a CMF of $15\pm7\%$, with ${\rm Fe/Mg}=0.84\pm0.28$ and ${\rm Fe/Si}=0.83^{+0.35}_{-0.25}$, about twice smaller than solar values, $3\,\sigma$ lower than FGK stars \citep[${\rm Fe/Mg}=1.98\pm0.30$ and ${\rm Fe/Si}=2.14\pm0.45$,][adopted from Table~4 in \citet{jahandar_chemical_2025} and converted from molar ratios]{brewer_spectral_2016}, and also quite far in the lower tail of the distribution for M dwarfs \citep[${\rm Fe/Mg}=1.59\pm0.60$ and ${\rm Fe/Si}=2.22\pm1.32$,][converted from molar ratios]{jahandar_chemical_2025}. Our estimation of the WMF in the solar (or stellar) prior case translates into a 3-$\sigma$ (2-$\sigma$) confidence that the planet has a non-zero amount of water. As shown in Figure~\ref{fig:cmf_wmf}, the uncertainty on the WMF is primarily driven by the constraint on the abundance ratio of refractory elements, particularly the Fe/Mg ratio. This highlights the critical need for empirical determination of these stellar abundance ratios through high-resolution spectroscopy.

The discussion in the next paragraph assumes that the WMFs we found are accurate, but we emphasize that certain assumptions were made to reach those results. Namely, the forward model assumes no water in the mantle or core, but \citet[][]{luo_interior_2024} showed that super-Earths (and sub-Neptunes) can store a majority of their water content in their mantle or core, and the assumed location of the water can affect the planetary radius by up to 15--25\%. We also apply solar (or stellar) priors on Fe/Mg and Fe/Si, but the planet may actually have different abundance ratios, in which case the priors we applied could bias the WMFs.

Keeping these disclaimers in mind and assuming the WMFs we found are accurate, this water content 
would be equivalent to several hundreds of TO ($322\pm138$\,TO with solar priors) which is at least an order of magnitude greater than the amount of water that could have been lost through XUV-driven evaporation over the lifetime of TRAPPIST-1. To cover the entire surface of TRAPPIST-1\,f with a 100-meter-deep ocean, in line with the definition of an ``aquaplanet'' from \citet{fauchez_impact_2019}, we only need about 0.04\,TO. Thus, assuming that TRAPPIST-1 and its planets have solar Fe/Mg and Fe/Si, TRAPPIST-1\,f may be a water world with its water content being stored in its core, mantle and/or on its surface. Among the atmospheric scenarios explored in \citet{fauchez_impact_2019}, assuming that TRAPPIST-1\,f is an aquaplanet (surface fully covered in water), all 1-bar, Archean Earth-like (N$_2$-dominated + CO$_2$ and CH$_4$) and \hbox{CO$_2$-dominated} atmospheres lead to a liquid body of water, either very localized near the substellar point or over the entire surface of the planet. We note however that a modern Earth-like atmosphere leads instead to a completely frozen surface due to a weaker greenhouse effect.

\section{Conclusion} \label{sec:conclusion}

We presented the first transit observations of TRAPPIST-1\,f with JWST NIRISS SOSS. At least one stellar flare was observed in each visit. 
Despite the amount of stellar flares, we found no evidence of stellar contamination from the TLSE in the transmission spectra obtained from two independent data reduction pipelines. If it is confirmed that TRAPPIST-1\,f is less affected by the TLSE than the closer-in planets of the system, this may imply a non-uniform latitudinal distribution of stellar surface heterogeneities. 
We reject cloud-free, H$_2$- and He-dominated atmospheres with surface pressures higher than about 3\,mbar at 95\% confidence, but cannot rule out the presence of a thick (\hbox{$>1\,$bar}) N$_2$-dominated atmosphere. Upper limits on the surface pressures of pure (single-species) atmospheres depend on whether we account for slopes in the transit spectra and on the reduction pipeline, despite the apparent similarity between the transit spectra, possibly due to the high sensitivity of retrievals to small differences in transit spectra and/or to the flares affecting (the light curve fit of) one reduction more than the other.

Looking forward and as demonstrated in this paper, SOSS observations of TRAPPIST-1\,f are not sensitive to CO$_2$ and can only place upper limits on a \textit{pure} CH$_4$ atmosphere. This adds onto previous works \citep[e.g.,][]{triaud_atmospheric_2023,trappist-1_jwst_community_initiative_roadmap_2024,fauchez_impact_2019} recommending observations of CO$_2$ features. For example, the 4.3\,$\mu$m band of an Archean Earth-like atmosphere with clouds and hazes would be detectable with NIRSpec at 5\,$\sigma$ with 14 transits \citep[assuming no TLSE or other types of stellar variability,][]{fauchez_impact_2019}. Such observations should be complemented with secondary eclipse observations, e.g., 15-$\mu$m eclipse photometry with MIRI would require 11 visits to achieve a 3-$\sigma$ detection of the eclipse depth for a dark (Bond albedo $A_B=0$), airless planet \citep{doyon_temperate_2024}. Conversely, the presence of an icy surface ($A_B \sim 0.3$) and/or a tenuous atmosphere with a small amount (e.g., $\sim$0.1\,ppm) of CO$_2$ would result in a significantly reduced, if not negligible, secondary eclipse depth at 15\,$\mu$m.


\section*{Note added in proof}

\citet{glidden_leveraging_2026} recently suggested that the moderate impact parameters of the outer TRAPPIST-1 planets may naturally mitigate the TLSE, consistent with our results for TRAPPIST-1. 

\begin{acknowledgments}
We thank the anonymous referee for comments that improved the quality of this manuscript. O.L. would like to thank Alexander D. Rathcke, Paul Charbonneau, Giovanni Bruno, Jozef Lipt\'{a}k, Rosa Keers, Elsa Ducrot, and Natalie Allen for insightful discussions that helped improve this work. The authors also wish to thank Kevin Volk for sharing the reference file required to flux-calibrate the SOSS spectra of TRAPPIST-1. 
The data presented in this paper were obtained from the Mikulski Archive for Space Telescopes (MAST) at the Space Telescope Science Institute. The specific observations analyzed can be accessed via \dataset[https://doi.org/10.17909/7tbe-2v53]{https://doi.org/10.17909/7tbe-2v53}. STScI is operated by the Association of Universities for Research in Astronomy, Inc., under NASA contract NAS5–26555. Support to MAST for these data is provided by the NASA Office of Space Science via grant NAG5–7584 and by other grants and contracts. 
These observations are associated with JWST GTO program \#1201. 
This project was undertaken with the financial support of the Canadian Space Agency. 
This research made use of the VizieR catalogue access tool, CDS, Strasbourg, France (DOI : 10.26093/cds/vizier), specifically, VizieR catalogue J/A+A/640/A112 (Ducrot E.) \citep{vizier_ducrot_trappist-1_2020}. The original description of the VizieR service was published in 2000, A\&AS 143, 23. 
O.L.\ acknowledges financial support from the Fonds de recherche du Qu\'{e}bec --- Nature et technologies (FRQNT) under file \#303926 (\url{https://doi.org/10.69777/303926}), and funding from the Trottier Family Foundation in their support of IREx. 
R.D.\ acknowledges financial support from the NSERC and the Canadian Space Agency through grants number 22JWG01-3 and 22EXPJWST.
A.L'H.\ acknowledges support from the FRQNT under file \#349961 (\url{https://doi.org/10.69777/349961}). 
C.P.-G.\ acknowledges support from the E. Margaret Burbidge Prize Postdoctoral Fellowship from the Brinson Foundation. 
M.T.\ acknowledges support from the Tremplin 2022 program of the Faculty of Science and Engineering of Sorbonne University. M.T. acknowledges support from BELSPO BRAIN (B2/212/PI/PORTAL). 
M.G.\ is F.R.S.-FNRS Research Director. M.G. and M.T. acknowledge support from the BELSPO program BRAIN-be 2.0 (Belgian Research Action through Interdisciplinary Networks), contract B2/212/B1/PORTAL. This work was supported by CNES, focused on the NIRISS instrument on JWST. 
L.D.\ is a Banting and Trottier Postdoctoral Fellow and acknowledges support from the Natural Sciences and Engineering Research Council (NSERC) and the Trottier Family Foundation. 
D.J.\ is supported by NRC Canada and by an NSERC Discovery Grant. 
S.P.\ acknowledges support from the Swiss National Science Foundation under grant 51NF40\_205606 within the framework of the National Centre of Competence in Research PlanetS. 
\end{acknowledgments}

%

\vspace{5mm}
\facilities{JWST(NIRISS)}


\software{\texttt{exoTEDRF}\footnote{\url{https://github.com/radicamc/exoTEDRF}} \citep{feinstein_early_2023,radica_applesoss_2022,radica_awesome_2023},
          \texttt{jwst}\footnote{\url{https://github.com/spacetelescope/jwst}},
          \texttt{emcee}\footnote{\url{https://emcee.readthedocs.io/en/stable/}} \citep{foreman-mackey_emcee_2013},
          \texttt{corner}\footnote{\url{https://corner.readthedocs.io/en/latest/}} \citep{foreman-mackey_cornerpy_2016},
          \texttt{batman}\footnote{\url{https://lkreidberg.github.io/batman/docs/html/index.html}} \citep{kreidberg_batman_2015},
          \texttt{celerite}\footnote{\url{https://celerite.readthedocs.io/en/stable/}} \citep{foreman-mackey_fast_2017},
          \texttt{POSEIDON}\footnote{\url{https://poseidon-retrievals.readthedocs.io/en/latest/}} \citep{macdonald_hd_2017,macdonald_poseidon_2023},
          \texttt{PandExo}\footnote{\url{https://exoctk.stsci.edu/pandexo/}} \citep{batalha_pandexo_2017},
          PHOENIX \citep{husser_new_2013},
          SPHINX \citep{iyer_sphinx_2023,iyer_sphinx_2023_7416042},
          \texttt{MSG}\footnote{\url{https://msg.readthedocs.io/en/stable/}} \citep{townsend_msg_2023},
          \texttt{PyMultiNest}\footnote{\url{https://johannesbuchner.github.io/PyMultiNest/}} \citep{Feroz2009,Buchner2014},
          \texttt{astropy}\footnote{\url{https://www.astropy.org/}} \citep{the_astropy_collaboration_astropy_2013,the_astropy_collaboration_astropy_2018},
          \texttt{numpy}\footnote{\url{https://numpy.org/}} \citep{harris_array_2020}, 
          \texttt{matplotlib}\footnote{\url{https://matplotlib.org/}} \citep{hunter_matplotlib_2007}, \texttt{scipy}\footnote{\url{https://scipy.org/}} \citep{virtanen_scipy_2020}
          }



\appendix

\section{Data reduction} \label{app:reduction}

\subsection{Asteroid crossing during Visits~1 and 3\label{app:reduction_asteroid}}

Prior to reducing the data, we noticed a light source resembling an order-0 field-star contaminant moving linearly on the spectral trace images over time in Visits~1 and 3. We identified these sources as asteroids. In Visit~1, the asteroid moved diagonally on the images, crossing a significant fraction of spectral order 2 but a smaller fraction of order 1. We found it simpler to keep the asteroid in the data all the way to the transit spectrum, where we masked the wavelength bins crossed by the asteroid. In Visit~3, the asteroid moved almost perpendicular to the spectral trace and therefore only a few pixels needed to be masked to remove its signal. These pixels were masked after the reduction, prior to the light curve fit. 



\subsection{Comparison between \texttt{SOSSISSE} and \texttt{exoTEDRF}}

\begin{figure*}
    \centering
    \includegraphics[width=\textwidth]{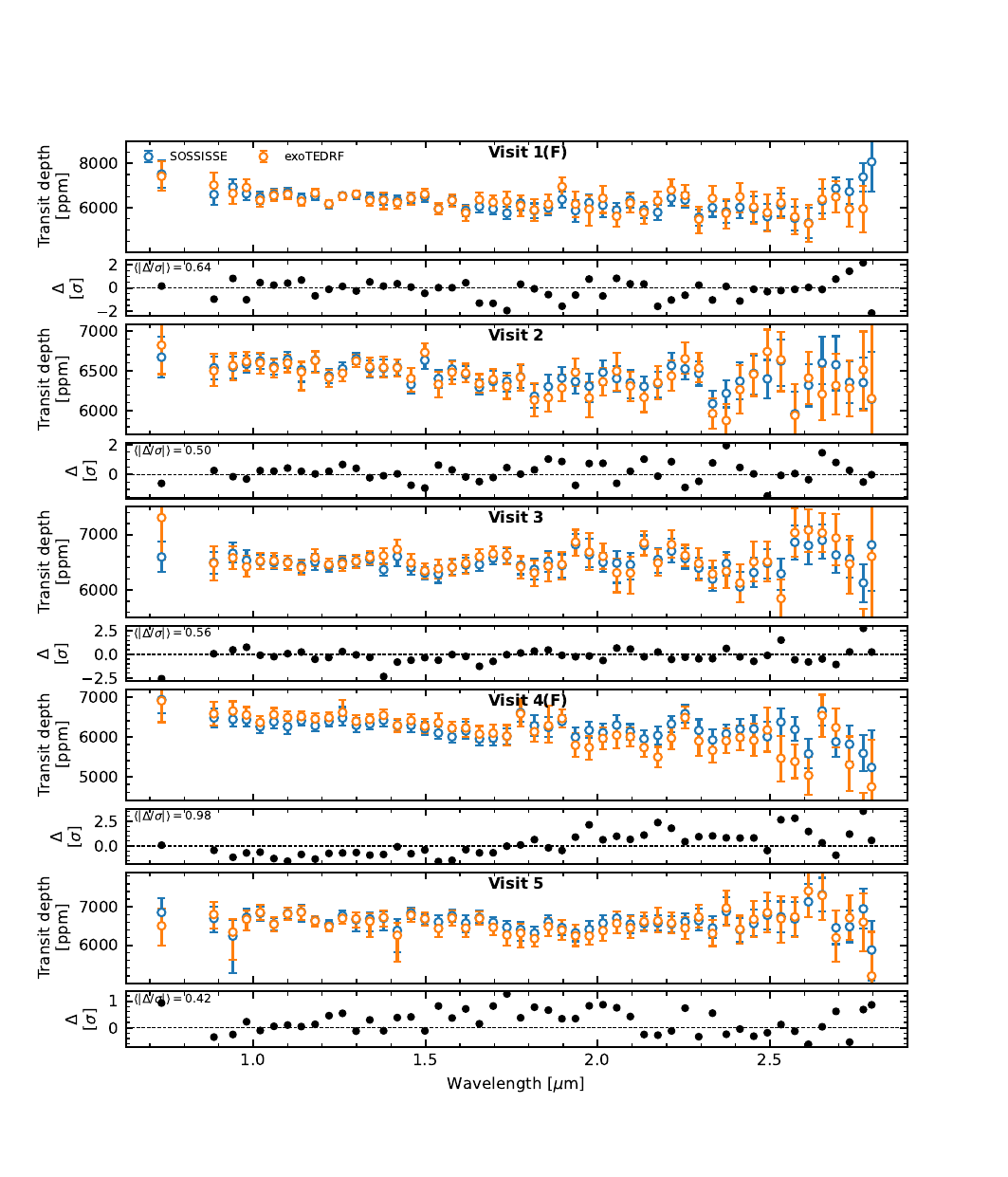}
    \caption{
    Comparison between the transit spectra from \texttt{SOSSISSE} (blue) and from \texttt{exoTEDRF} (orange). \texttt{exoTEDRF} spectra are vertically offset by $154$, $-19$, $-7$, $81$, and $-26$\,ppm 
    for Visits~1 to 5, respectively. Panels under each transit spectrum (black) show the differences between the two spectra, normalized by the smallest error bar between the two points in each bin. The mean normalized absolute difference is given in the top left corner of each panel. 
    (F) indicates visits with a flare near ingress.
    }
    \label{fig:tspec_pipelines}
\end{figure*}

Figure~\ref{fig:tspec_pipelines} compares the transit spectra from the \texttt{SOSSISSE} \citep{lim_atmospheric_2023,cadieux_transmission_2024} and the \texttt{exoTEDRF} \citep{feinstein_early_2023,coulombe_broadband_2023,radica_awesome_2023,radica_exotedrf_2024,radica_promise_2024} reduction pipelines for the five TRAPPIST-1\,f visits. In the \texttt{exoTEDRF} pipeline, the $1/f$ noise correction was performed at the integration level (instead of the group level), as is done in \texttt{SOSSISSE}, and we used the achromatic scaling method to scale the median stack \citep{radica_exotedrf_2024}. The light curves of the two reductions were obtained using the same priors, systematics models, and MCMC setup. The same orbital and limb darkening parameter fixing procedure (Section~\ref{sec:lcfit}) was applied to both reductions, but we allowed each reduction to find their own optimal parameter values. In Figure~\ref{fig:tspec_pipelines}, the transit spectra from \texttt{exoTEDRF} are vertically offset to match the \texttt{SOSSISSE} spectra. 
For each visit, we computed the differences between the two transit spectra and normalized them by the smallest uncertainty between the two points in each bin. The mean of these normalized (absolute) differences, $\left<|\Delta/\sigma|\right>$, is smaller than 1 in all visit, meaning that on average, differences between the transit spectra from the two reductions are well within the uncertainties.

The transit spectra that differ the most from one reduction to the other are those from Visits~1 and 4 ($\left<|\Delta/\sigma|\right>=0.64$ and $0.98$, respectively), the two visits most affected by flares. It is unclear whether flares affect the \textit{reduction}, but they can certainly affect the \textit{light curve fit}, and they may do so differently when the initial extracted spectra are slightly different. We note however that despite these differences, the mean normalized absolute differences remain lower than 1 in all five visits. In other words, on average, the difference between the two reductions in a given spectral bin is smaller that the smallest error bar between these two points.

\section{Light curve fit} \label{app:lcfit}

\subsection{Parameter fixing} \label{app:lcfit_paramfix}

We fixed orbital parameters (impact parameter $b$ and normalized semi-major axis $a/R_\star$) and limb darkening coefficients (LDCs) to reduce the degeneracy between these parameters and $R_p/R_s$, and thus reduce the uncertainties on the transit spectra. Orbital parameters should be the same across all TRAPPIST-1\,f visits and LDCs should be the same across all TRAPPIST-1 visits because the same planet and star are observed, and a good value to adopt is one that contains information from multiple state-of-the-art, high-SNR light curves from JWST. For the LDCs, one could argue that stellar surface heterogeneities and stellar variability could impact the LDCs, but by fixing them, we force the \textit{systematics models} (linear function, GP, H$\alpha$ detrending) to account for this stellar heterogeneity/variability, as they are meant to do. 

With observations that are contaminated by stellar flares, we believe it is particularly important to provide prior information on the LDCs to guide the MCMC walkers towards the most likely values based on what we know of the star, and we have such information coming from previous TRAPPIST-1 observations. While it is true that in previous works on TRAPPIST-1 planets with JWST data, the LDCs were fitted \citep[e.g., for TRAPPIST-1\,b and c with NIRISS/SOSS, the \citet{kipping_efficient_2013} parameterization was used with uniform priors][]{lim_atmospheric_2023,radica_promise_2024}, there is precedent of fixing the LDCs to values based on previous observations, especially when the observations we try to analyze are contaminated by stellar flares (which is the case for at least three of the five visits presented here): for two flare-contaminated visits of TRAPPIST-1\,d with NIRSpec/Prism, the LDCs were fixed to values from an observation of TRAPPIST-1\,g which occurred when the star was relatively less active \citep{piaulet-ghorayeb_strict_2025}.

\begin{figure*}
    \centering
    \includegraphics[width=\textwidth]{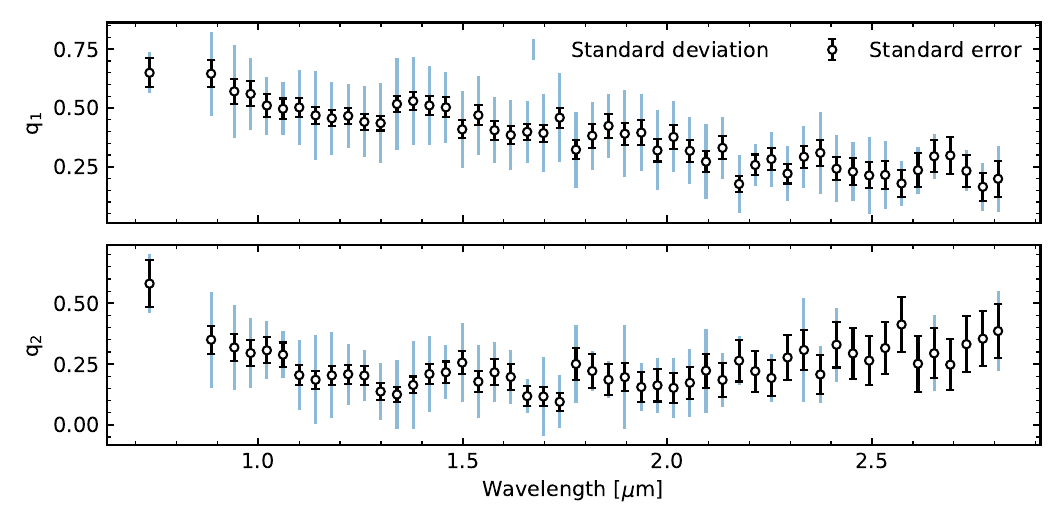}
    \caption{Limb darkening coefficients $q_1$ (top) and $q_2$ (bottom) computed by taking the per-bin error-weighted average of all existing TRAPPIST-1 observations with JWST/NIRISS/SOSS \citep[TRAPPIST-1\,b and c from][respectively, and TRAPPIST-1\,f from this work]{lim_atmospheric_2023,radica_promise_2024}. Black error bars show the standard error on the weighted averages while blue bars show the standard deviation across observations.}
    \label{fig:ldcs_spread}
\end{figure*}

To fix the orbital parameters and LDCs to optimal values, we fitted the light curves iteratively in four steps for order 1 and in two steps for order 2. 
For order 1, the first step consists in fitting the broadband light curve of each visit independently, with the parameters listed with priors in Table~\ref{tab:priors_posteriors_lcfit} as free parameters. We then computed the error-weighted average of all available (SOSS order 1 broadband) LDCs of TRAPPIST-1 \citep{lim_atmospheric_2023,radica_promise_2024}, including those of the five TRAPPIST-1\,f visits. 
The second step consists in fitting the broadband light curve of each visit again, now fixing the LDCs to the average computed in the first step (top and middle panels of each visit in Figure~\ref{fig:lcfit}). We then computed the error-weighted average of the impact parameter and semi-major axis across the five visits of TRAPPIST-1\,f. 
The third step consists in fitting the spectroscopic light curves of each visit, fixing the impact parameter and semi-major axis to the average computed in the second step (the mid-transit times were fixed to the value obtained in the second step for each visit), but with the (spectroscopic) LDCs as free parameters. The priors of all the free parameters are the same as in the broadband light curve fit. We then computed the per-bin, error-weighted averages of all available (SOSS order 1 spectroscopic) LDCs of TRAPPIST-1 (Figure~\ref{fig:ldcs_spread}). 
The fourth and last step consists in fitting the spectroscopic light curves of each visit, now additionally fixing the LDCs to the averages computed in the third step. 
For order 2, the first step consists in fitting the light curve of each visit, fixing the impact parameter, semi-major axis, and mid-transit times to the same values as in the third step of order 1. We then computed the error-weighted average of all available (SOSS order 2 broadband) LDCs of TRAPPIST-1. 
The second and last step consists in fitting the light curve of each visit, now additionally fixing the LDCs to the average computed in the first step.

\begin{figure*}
    \centering
    \includegraphics[width=\textwidth]{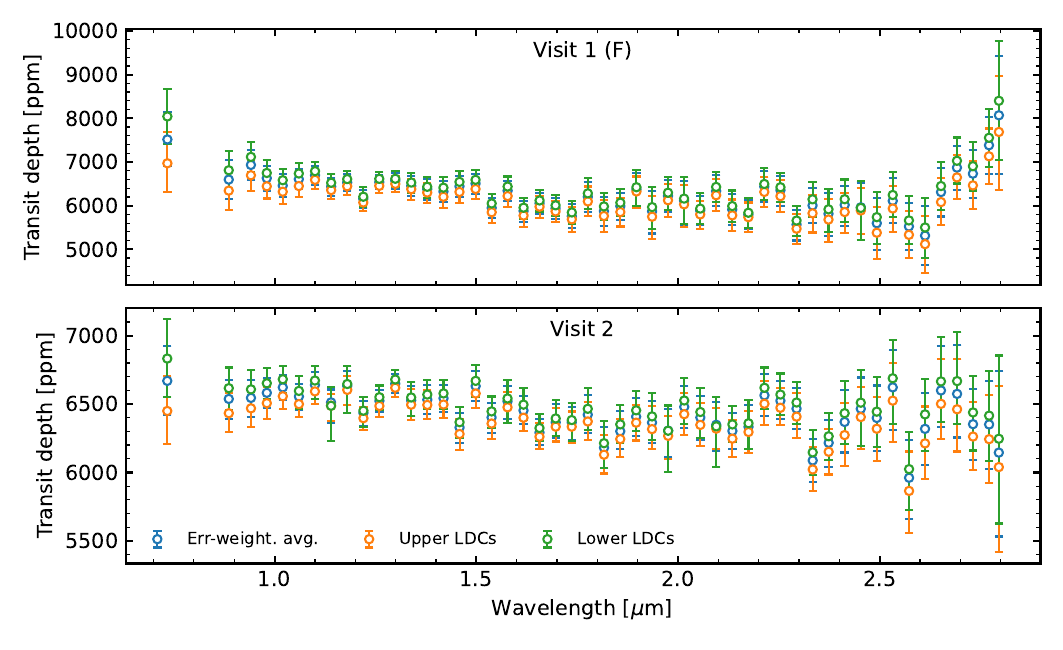}
    \caption{Transit spectrum of Visits~1 (top) and 2 (bottom) obtained by fixing the limb darkening coefficients (LDCs) to three different sets of values: 1) error-weighted average LDCs (blue, i.e., identical to the transit spectra in panel a of Figure~\ref{fig:tspec_retrieval-posteriors}), 2) error-weighted average LDCs \textit{plus} their standard error (orange), and 3) error-weighted average LDCs \textit{minus} their standard error (green). As a reminder, in Visit~1, a high-energy flare occurred shortly before the transit ingress while in Visit~2, a lower-energy flare occurred long before the ingress. We chose to present Visits~1 and 2 only instead of all five visits for simplicity.}
    \label{fig:tspec_upper-lower-ldcs}
\end{figure*}

To assess the impact of fixing the LDCs, we focused on two representative visits: Visit~1, in which a relatively high-energy flare occurred shortly before the transit ingress, and Visit~2, in which a lower-energy flare occurred long before the ingress. For each of these two visits, we fitted the spectroscopic light curves by fixing the LDCs to two additional sets of LDCs (in addition to the aforementioned, nominal, error-weighted averaged LDCs): 1) the error-weighted average LDCs to which we added the standard error on the (error-weighted average) LDCs and 2) the error-weighted averaged LDCs from which we subtracted the standard error (Figure~\ref{fig:tspec_upper-lower-ldcs}). These two sets of LDCs can be interpreted as the upper- and lower-bounds of the error-weighted average LDCs, and the transit spectra obtained by fixing the LDCs to these bounds provide an estimate of the transit depth uncertainty that is unaccounted for by fixing the LDCs. Figure~\ref{fig:tspec_upper-lower-ldcs} shows that in all wavelength bins, the transit depths obtained with the upper- and lower-bound LDCs are within the uncertainty of the transit depths obtained with the nominal, error-weighted average LDCs. We do acknowledge that fixing the LDCs may artificially shrink the transit depth uncertainty, especially in spectral order 2. Since order 2 \textit{may} drive the detection of the TLSE effect, we warn that our non-detection of the TLSE may not be as confident if we account for the uncertainty on the LDCs.

In both the broadband and spectroscopic light curve fits, when the limb darkening coefficients were fitted, we adopted the quadratic limb darkening law with the parameterization from \citet{kipping_efficient_2013}. We also tested the power2 limb darkening law \citep{hestroffer_centre_1997}, but saw no significant impact on the transit spectrum.

We provide the LDCs of each visit in Zenodo via\dataset[DOI: 10.5281/zenodo.14501417]{https://doi.org/10.5281/zenodo.14501417} for future works.

\subsection{Systematics models} \label{app:lcfit_sysmodels}

The SOSS light curves of TRAPPIST-1\,b presented in \citet{lim_atmospheric_2023} were fitted with a linear function, and either a detrending against the H$\alpha$ flux, a squared-exponential-kernel Gaussian process (GP), both a detrending against H$\alpha$ and a GP, or a GP that was first trained on the H$\alpha$ time series. It was shown that the differences in the transit spectra resulting from these different systematics treatments were generally small compared to the transit depth error bars (their Figure~6). Here we explain how we selected the systematics treatments to be applied in the light curve fit presented in Section~\ref{sec:lcfit}. 

We started from the assumption that a simple linear function is not sufficient to remove the observed stellar variability, but is essential to capture any possible long-term trends. We then compared the transit spectra resulting from light curve fits that included either one of two GP kernels: the squared-exponential (SE) kernel and the simple-harmonic-oscillator (SHO) kernel. We found little to no differences in the transit spectra, but found that the light curve fit residuals showed less correlated noise with the SHO kernel. We thus opted for the SHO kernel. 

\begin{figure*}
    \centering
    \includegraphics[width=\textwidth]{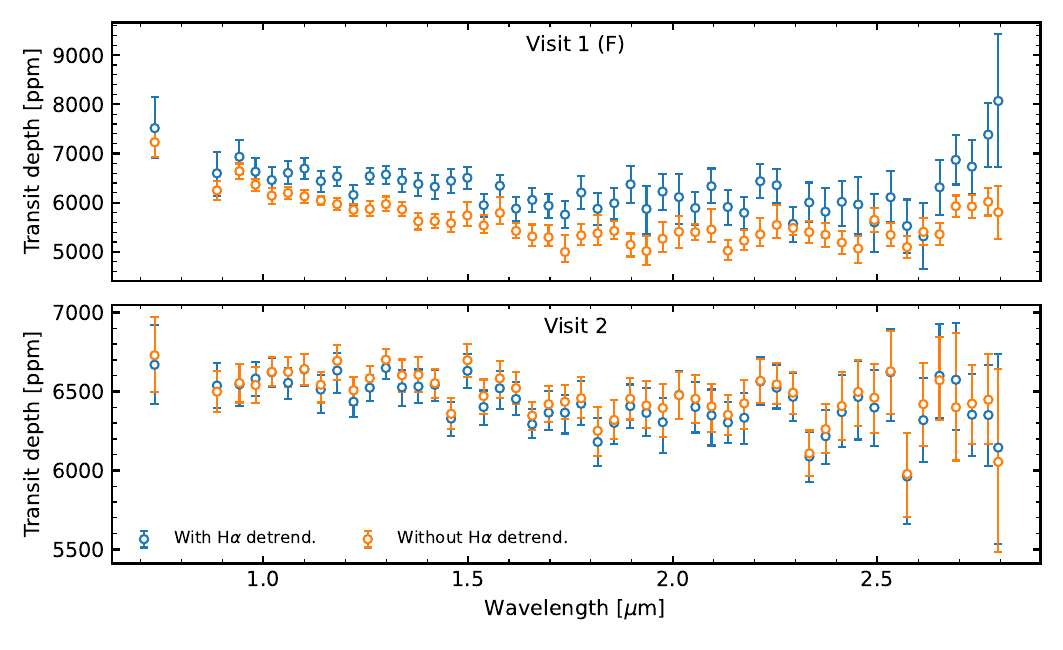}
    \caption{Transit spectrum of Visits~1 (top) and 2 (bottom) obtained by detrending the light curves against the H$\alpha$ timeseries (blue) or not (orange). All other parameters of the light curve fitting remain the same. In particular, a linear function and a Gaussian process was fitted simultaneously with the transit model in all cases.}
    \label{fig:tspec_with-out-Ha}
\end{figure*}

Since at least two of the five transits were severely affected by a stellar flare, we suspected that a GP was insufficient to capture the short-timescale signature of the flares, so we detrended the broadband and spectroscopic light curves against the H$\alpha$ flux timeseries. Whether we detrended against H$\alpha$ or not had little impact on the transit spectra of Visits~2, 3, and 5, but it had a significant impact of the transit spectra of Visits~1 and 4, the two visits most affected by flares (Figure~\ref{fig:tspec_with-out-Ha}). In Visit~1, detrending against H$\alpha$ produced the increase at red wavelengths (Section~\ref{sec:tspec}), leaving the rest of the spectrum relatively flat, while not detrending against H$\alpha$ produced a $1000-1500$\,ppm increase bluewards of $\sim$$1.7\,\mu$m with a smaller ($\sim$$500$-ppm) increase at red wavelengths, thus producing a quadratic-function-like spectrum. In Visit~4, not detrending against H$\alpha$ led to a larger slope in the entire spectrum ($-295^{+75}_{-76}\,$ppm/$\mu$m compared to $-208^{+61}_{-60}\,$ppm/$\mu$m when detrending against H$\alpha$). These increases and slopes are unlikely to be planetary in origin given their amplitudes, so we settled on the option that minimizes such signatures, which is to detrend against H$\alpha$. The fact that the H$\alpha$ detrending mostly affected the transit spectrum of flare-contaminated visits suggests that it is indeed removing flare signal and is thus a relevant, albeit imperfect, flare correction model. 

\begin{figure*}
    \centering
    \includegraphics[width=.3\textwidth]{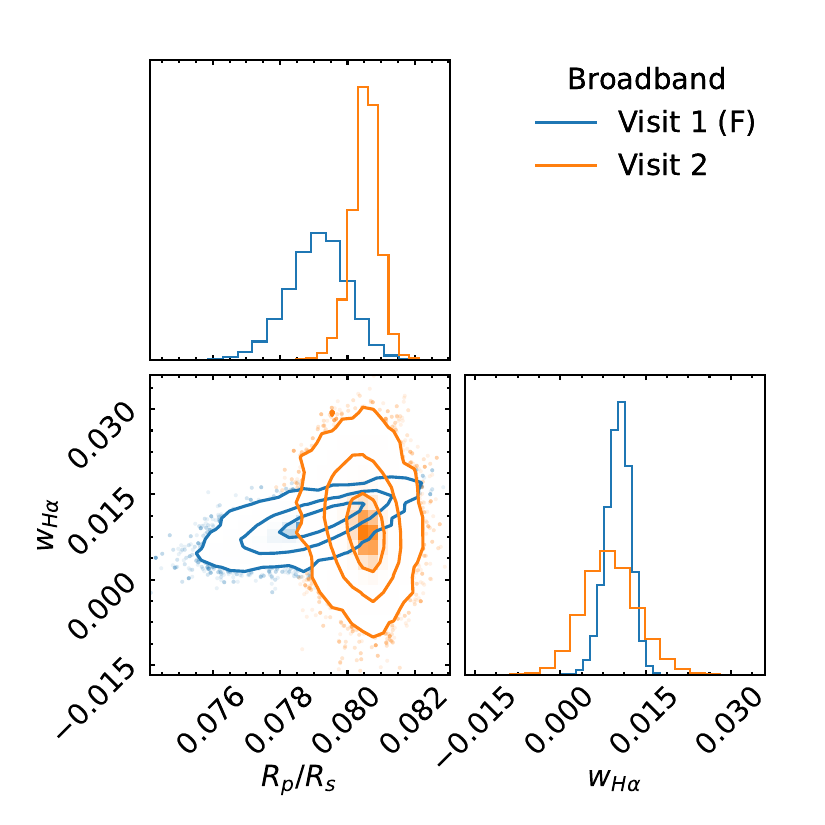}
    \includegraphics[width=.3\textwidth]{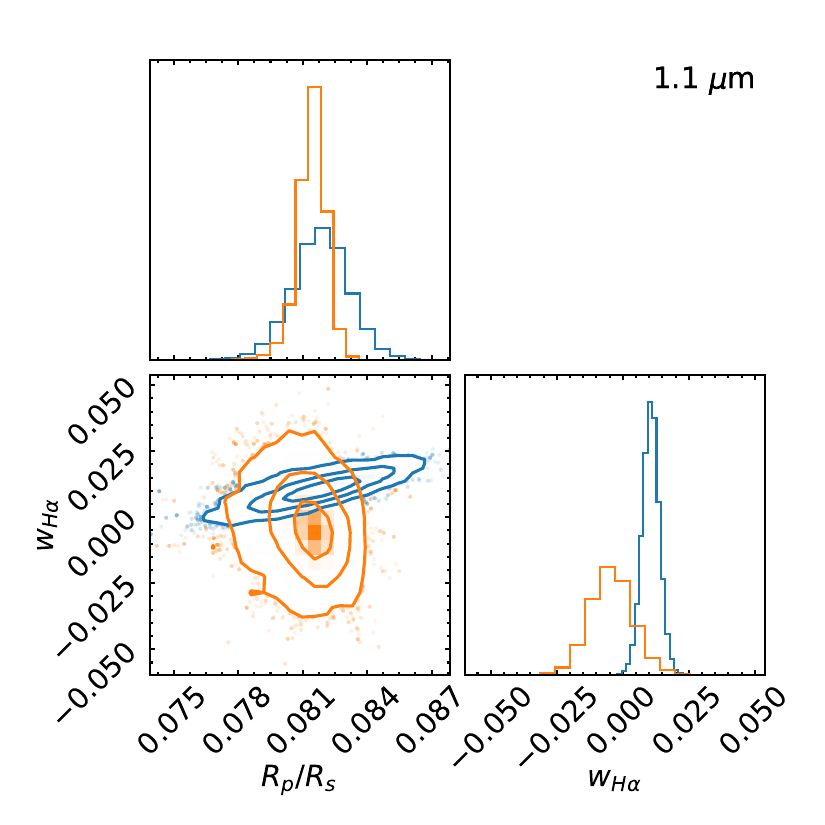}
    \includegraphics[width=.3\textwidth]{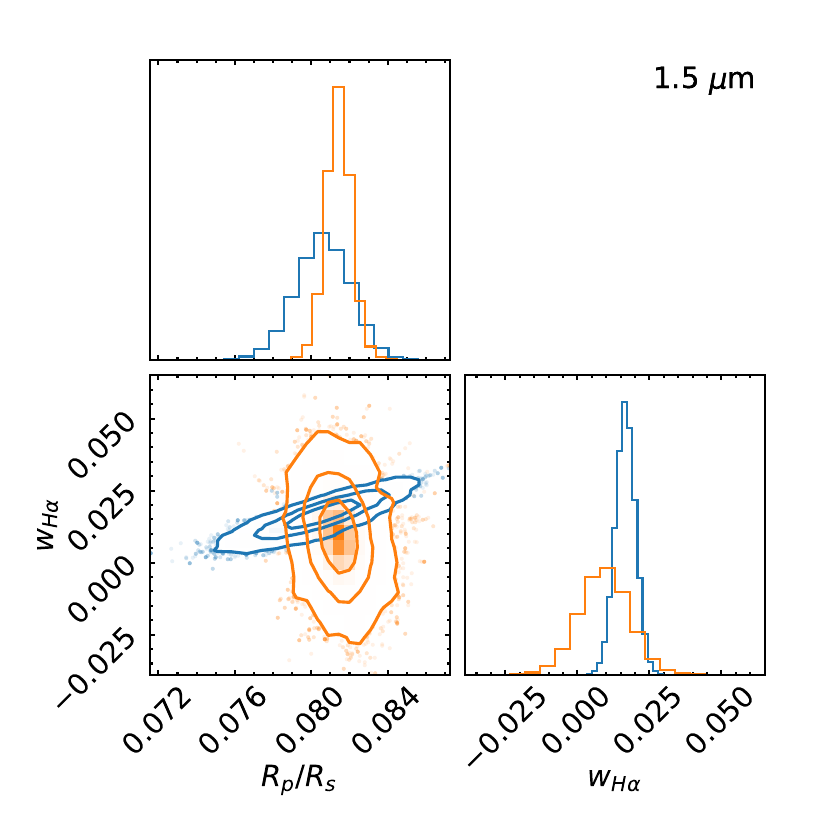}
    \includegraphics[width=.3\textwidth]{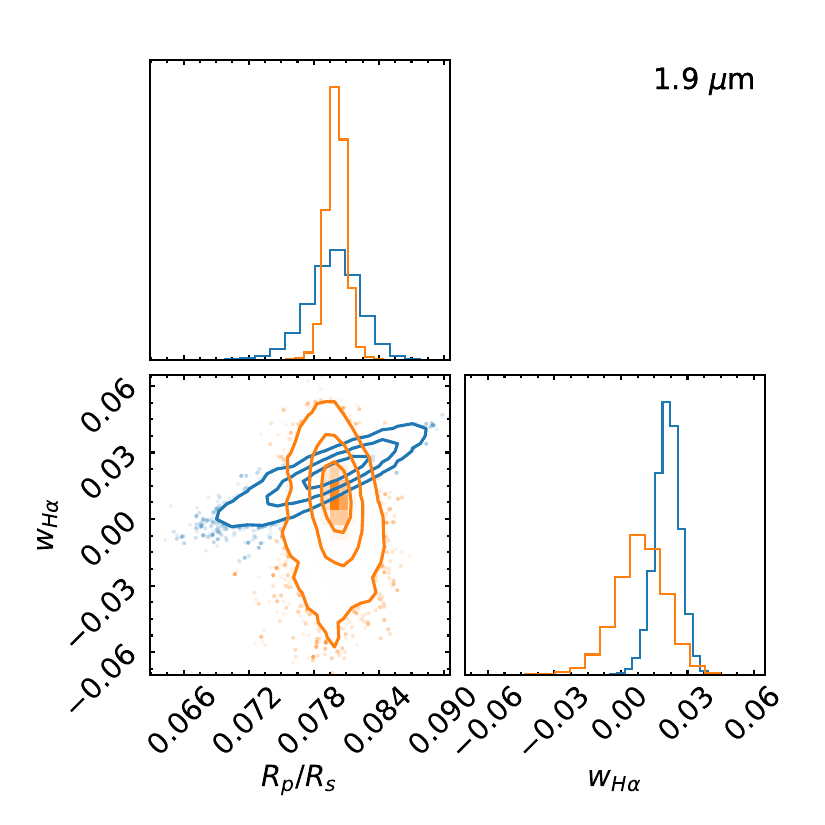}
    \includegraphics[width=.3\textwidth]{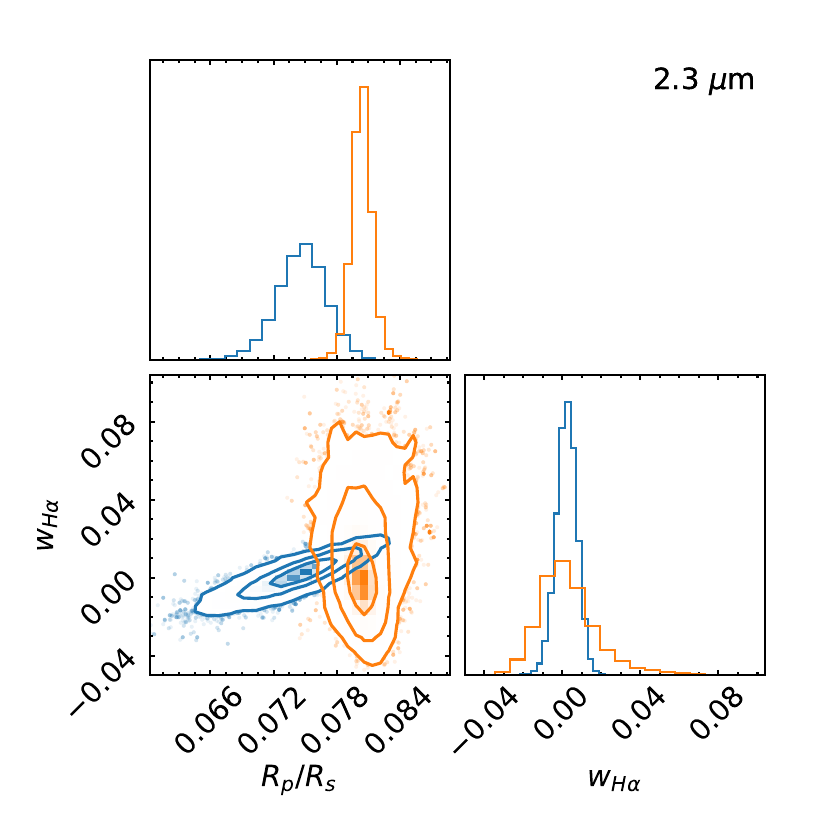}
    \includegraphics[width=.3\textwidth]{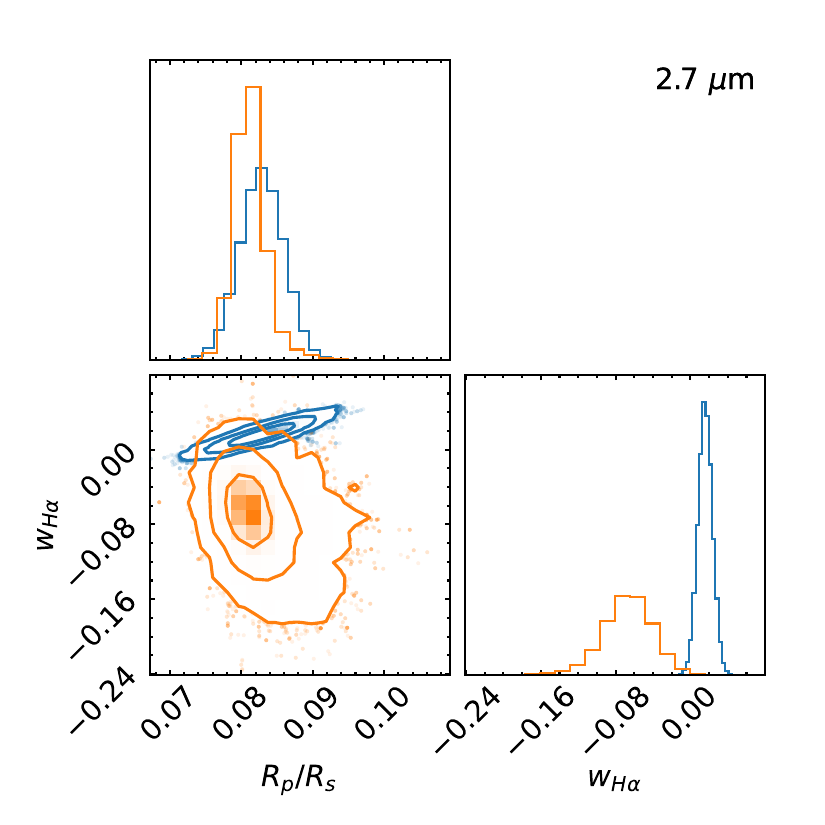}
    \caption{Joint and marginalized posterior distributions of the planet-to-star radius ratio $R_p/R_s$ and the free parameter $w_{{\rm H}\alpha}$ used to detrend the light curves against the H$\alpha$ time series. Top left corner plot shows the posteriors for the broadband light curve fit. Other corner plots are for five different wavelength bins, selected to roughly span the SOSS domain at equal wavelength intervals. In all corner plots, blue posteriors are for Visit~1, orange for Visit~2.}
    \label{fig:corner_Ha_vis-1-2}
\end{figure*}

We also do not have any strong arguments against the H$\alpha$ detrending: if it turns out that the H$\alpha$ time series does not correlate well with any structure in the light curve, the free multiplicative factor \citep[$w$ in Equation~B2 of][and $w_{{\rm H}\alpha}$ in Table~\ref{tab:priors_posteriors_lcfit}]{lim_atmospheric_2023} is allowed to go to zero. Figure~\ref{fig:corner_Ha_vis-1-2} shows the joint posteriors of $R_p/R_s$ and $w_{{\rm H}\alpha}$ for Visits~1 and 2 (Visit~1 being strongly flare contaminated and Visit~2, less so) for the broadband light curve fit and a selection of spectroscopic light curve fits. The posteriors show that the two parameters are more correlated in Visit~1 than in Visit~2. This is an expected behavior because we expect the flare signal in Visit~1 to dilute the transit, and therefore the H$\alpha$ signal should be correlated with the transit depth (e.g., by multiplying the H$\alpha$ time series with a larger factor, the transit model has to compensate this large dilution by increasing $R_p/R_s$). For Visit~2, the H$\alpha$ time series is essentially noise during the transit (see panel e of Figure~\ref{fig:lcfit}), and multiplying it by a larger or smaller factor has no effect on the transit depth. The correlation between the two parameters in Visit~1 \textit{should} be accounted for when estimating the uncertainty on $R_p/R_s$, and we do so by including the H$\alpha$ detrending in the fit. \textit{Not} detrending against H$\alpha$ would be equivalent to fixing $w_{{\rm H}\alpha}$ to zero and not accounting for its correlation with $R_p/R_s$, thus underestimating the transit depth uncertainty. The $w_{{\rm H}\alpha}$ posteriors are also narrower and generally less consistent with zero in Visit~1 than in Visit~2, suggesting that this detrending is mostly relevant in the flare-contaminated Visit~1, but is not affecting the flare-uncontaminated Visit~2 given the convergence towards values close to zero. Figure~\ref{fig:corner_Ha_vis-1-2} and this discussion only focus on Visits~1 and 2, but the same conclusions can be drawn from the other three visits. The only exception is that in Visit~4, $R_p/R_s$ and $w_{{\rm H}\alpha}$ are less correlated than in Visit~1, but this could be explained by the fact that the H$\alpha$ line only emits significantly during the first half of the transit in Visit~4 whereas it emits strongly during the entire transit of Visit~1 (compare panels l and c in Figure~\ref{fig:lcfit}).

\begin{figure*}
    \centering
    \includegraphics[width=.32\textwidth]{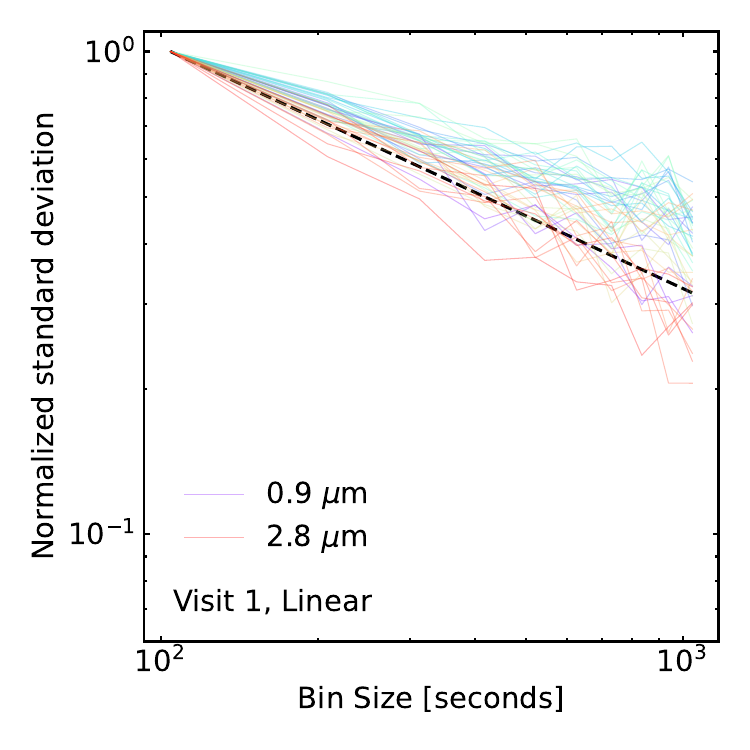}
    \includegraphics[width=.32\textwidth]{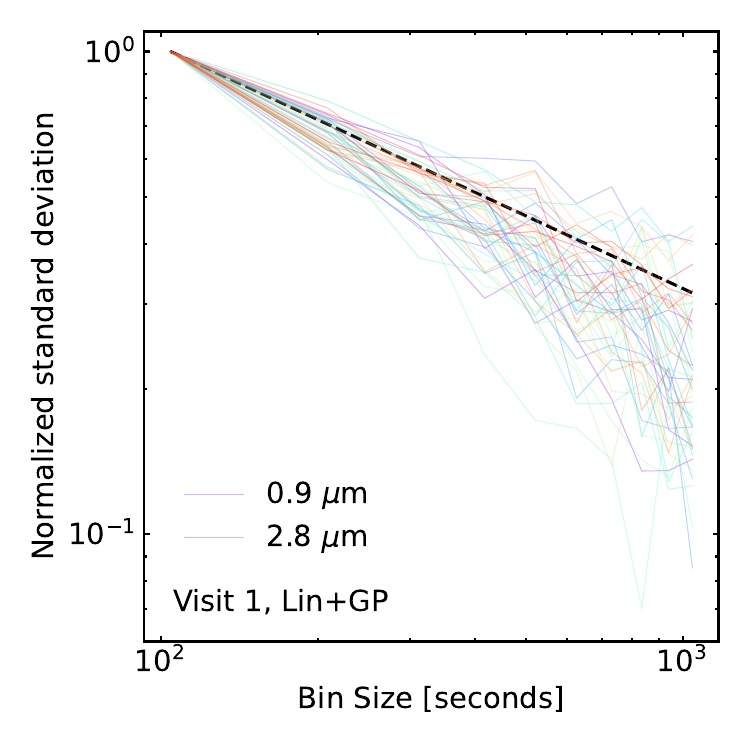}
    \includegraphics[width=.32\textwidth]{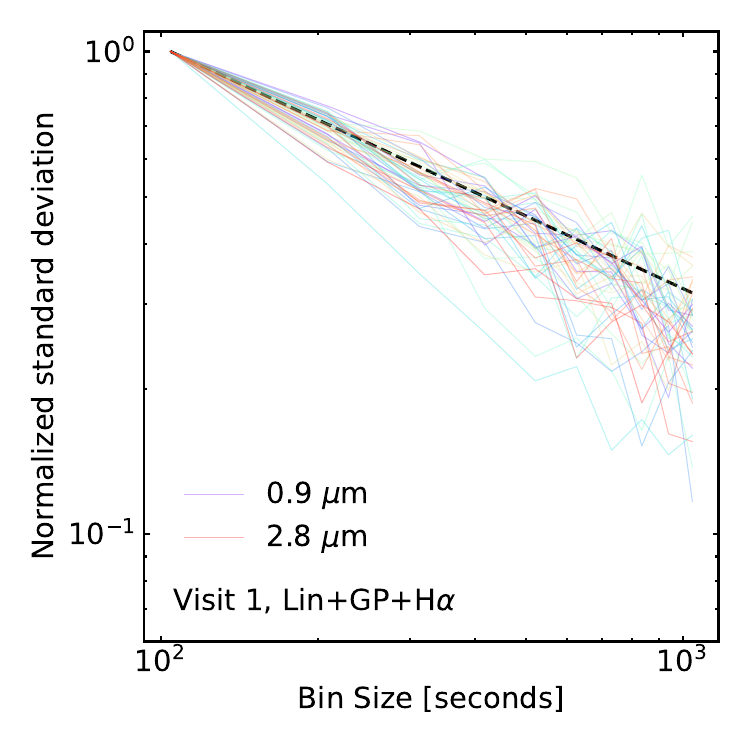}
    \caption{
    Normalized standard deviation of the residuals from the light curve fits of Visit~1 as a function of bin size. The three panels correspond to three levels of systematics correction: linear function only (left); linear function and Gaussian process (middle); and linear function, Gaussian process, and H$\alpha$ detrending (right). Different colors correspond to different wavelength bins, the most violet being the shortest wavelength and the reddest, the longest.
    }
    \label{fig:binned_std}
\end{figure*}

\begin{figure*}
    \centering
    \includegraphics[width=.32\textwidth]{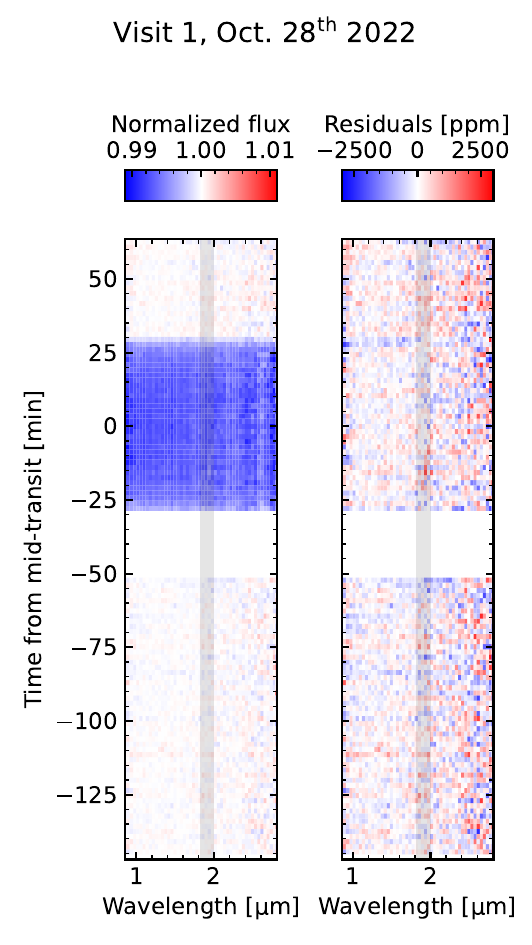}
    \includegraphics[width=.32\textwidth]{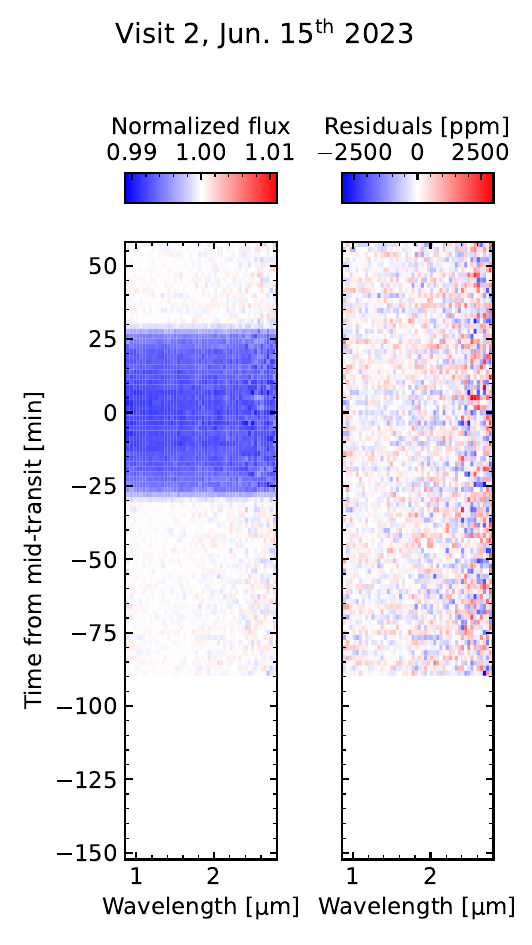}
    \includegraphics[width=.32\textwidth]{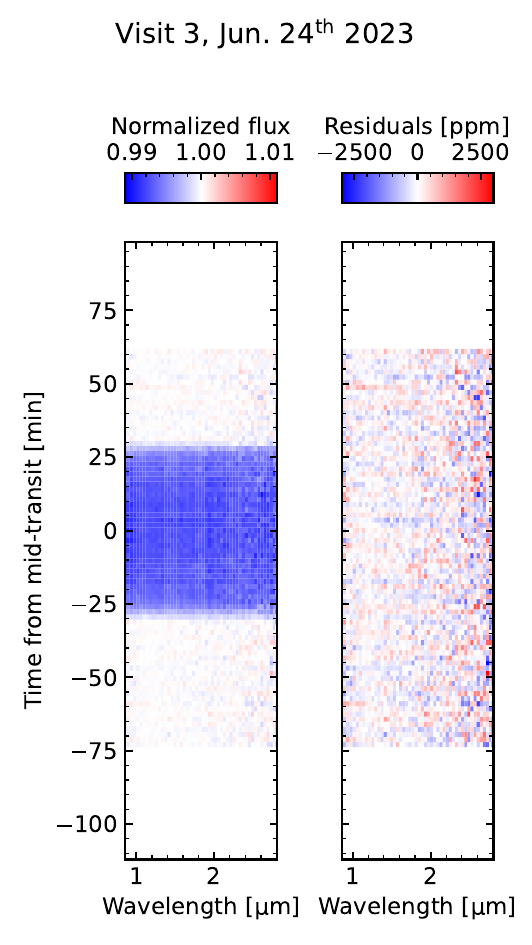}
    \includegraphics[width=.32\textwidth]{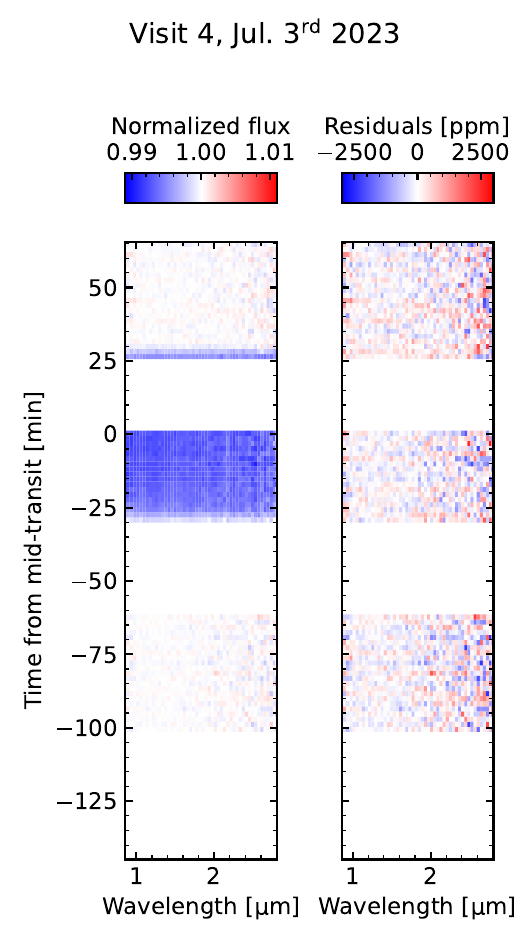}
    \includegraphics[width=.32\textwidth]{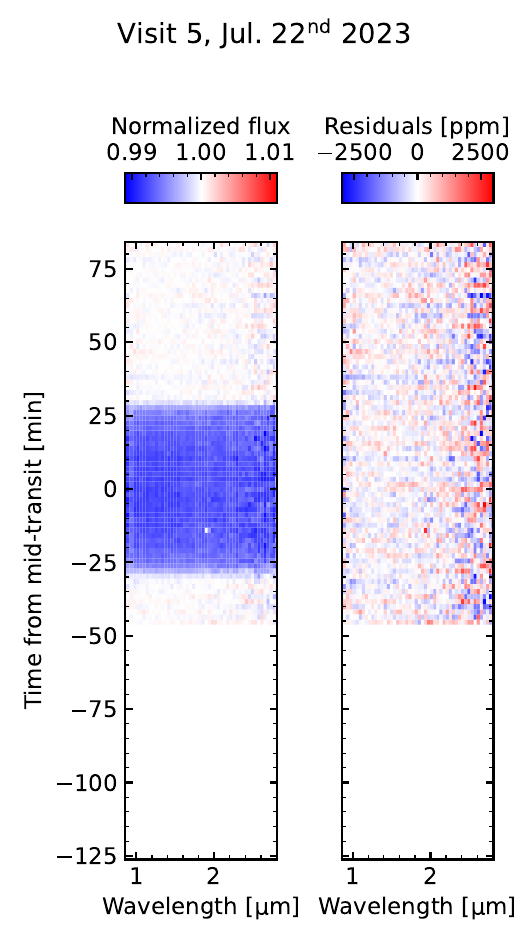}
    \caption{
    Systematics-corrected, spectroscopic light curves (left of each pair of panels) and residuals from the best-fitting transit and systematics model (right of each pair of panels). Panels are displayed from left to right and from top to bottom in chronological order of observation. White spaces represent data that were masked during the light curve fit. Shaded regions in Visit~1 indicate the wavelengths that the asteroid crossed on the spectral trace.
    }
    \label{fig:slcs_residuals}
\end{figure*}

Figure~\ref{fig:binned_std} shows the normalized standard deviation of the residuals from the light curve fits of Visit~1 as a function of bin size for three different levels of systematics correction to assess over-/under-correction. 
Figure~\ref{fig:slcs_residuals} shows, for each of the five visits, the (systematics-corrected) spectroscopic light curves and residuals from the best-fitting transit and systematics models at the end of the iterative fitting procedure.

\begin{deluxetable*}{lcccccc}
        \tabletypesize{\scriptsize}
        \tablewidth{0pt} 
        \tablecaption{Priors and posteriors of the broadband light curve fit parameters and derived parameters.}
        \tablehead{
        \colhead{Parameter} & \colhead{Prior} & \multicolumn{5}{c}{Posterior}
        } 
        \startdata
         & & Visit~1 (F) & Visit~2 & Visit~3 & Visit~4 (F) & Visit~5 \\
        \hline
        \multicolumn{7}{l}{Visit-specific parameters} \\
        \\
        $T_{\rm mid}$ (${\rm BJD}-2460000$) & $\mathcal{U}(-0.1, +0.1) + T_{\rm A21}$ & $-118.964953_{-0.000074}^{+0.000083}$ & $111.185737_{-0.000049}^{+0.000050}$ & $120.387900_{-0.000054}^{+0.000058}$ & $129.593860_{-0.000066}^{+0.000066}$ & $148.005161_{-0.000059}^{+0.000059}$ \\
        $(R_p/R_s)^2$ (ppm) & $\mathcal{U}(100, 50000)$ & $6261_{-141}^{+131}$ & $6485.3_{-66.7}^{+57.9}$ & $6496.0_{-75.8}^{+68.5}$ & $6234_{-137}^{+130}$ & $6633_{-171}^{+103}$ \\
        $s$ (ppm) & $\mathcal{U}(0, 2\times\textrm{estimated scatter})$ & $152.6_{-11.6}^{+12.7}$ & $144.8_{-11.0}^{+12.9}$ & $153.1_{-14.0}^{+14.7}$ & $131.6_{-12.6}^{+14.4}$ & $163.5_{-13.9}^{+16.3}$ \\
        $v$ ($10^{-6}$ days$^{-1}$) & $\mathcal{U}(-10^{10}, 10^{10})$ & $5012_{-952}^{+872}$ & $284_{-1403}^{+1110}$ & $1396_{-1342}^{+1347}$ & $6_{-2648}^{+2041}$ & $-8758_{-2916}^{+3225}$ \\
        $c$ & $\mathcal{U}(0.9, 1.1)$ & $0.999873_{-0.000088}^{+0.000080}$ & $0.999805_{-0.000075}^{+0.000076}$ & $0.999845_{-0.000091}^{+0.000101}$ & $0.99978_{-0.00014}^{+0.00016}$ & $0.99927_{-0.00012}^{+0.00012}$ \\
        $w_{\rm H\alpha}$ & $\mathcal{U}(-5, 5)$ & $0.0103_{-0.0022}^{+0.0021}$ & $0.0087_{-0.0046}^{+0.0051}$ & $0.0056_{-0.0060}^{+0.0048}$ & $0.0153_{-0.0022}^{+0.0023}$ & $0.0042_{-0.0037}^{+0.0033}$ \\
        $\log_{10}(a_{\rm GP})$ & $\mathcal{U}(1.75, 4)$ & $2.14_{-0.19}^{+0.20}$ & $2.00_{-0.17}^{+0.24}$ & $1.98_{-0.16}^{+0.25}$ & $2.39_{-0.19}^{+0.21}$ & $2.15_{-0.26}^{+0.38}$ \\
        $\log_{10}(\lambda_{\rm GP} ({\rm min}))$ & $\mathcal{U}(0.9, 2.2)$ & $1.86_{-0.38}^{+0.22}$ & $1.84_{-0.40}^{+0.25}$ & $1.75_{-0.65}^{+0.35}$ & $1.96_{-0.23}^{+0.17}$ & $1.85_{-0.53}^{+0.27}$ \\
        \\
        \hline
        \multicolumn{7}{l}{Parameters fitted for each visit, then fixed to a weighted average} \\
        \\
        $q_{1, \rm ord. 1}$ & $\mathcal{U}(0, 1)$ & \multicolumn{5}{c}{$0.400\pm0.050$} \\
        $q_{2, \rm ord. 1}$ & $\mathcal{U}(0, 1)$ & \multicolumn{5}{c}{$0.247\pm0.048$} \\
        $q_{1, \rm ord. 2}$ & $\mathcal{U}(0, 1)$ & \multicolumn{5}{c}{$0.650\pm0.088$} \\
        $q_{2, \rm ord. 2}$ & $\mathcal{U}(0, 1)$ & \multicolumn{5}{c}{$0.581\pm0.097$} \\
        $b$ & $\mathcal{N}(0.312, 0.023^2)$ & \multicolumn{5}{c}{$0.313\pm0.010$} \\
        $a/R_\star$ & $\mathcal{U}(35, 105)$ & \multicolumn{5}{c}{$69.48\pm0.27$} \\
        \hline
        \multicolumn{7}{l}{Derived parameters} \\
        \\
        $(R_p/R_s)$ & --- & $0.07913_{-0.00089}^{+0.00082}$ & $0.08053_{-0.00042}^{+0.00036}$ & $0.08060_{-0.00047}^{+0.00042}$ & $0.07896_{-0.00087}^{+0.00082}$ & $0.0814_{-0.0011}^{+0.0006}$ \\
        $(R_p/R_s)$ (weighted avg.) & --- & \multicolumn{5}{c}{$0.08032\pm0.00027$} \\
        $\rho_\star$ (g\,cm$^{-3}$) & --- & \multicolumn{5}{c}{$74.85\pm0.87$} \\
        $\rho_\star/\rho_\odot$ & --- & \multicolumn{5}{c}{$53.16\pm0.62$} \\
        $R_\star/R_\odot$ & --- & \multicolumn{5}{c}{$0.1191\pm0.0011$} \\
        $R_p$ ($R_\oplus$) & --- & \multicolumn{5}{c}{$1.043\pm0.010$} \\
        \\
        \enddata
        \tablecomments{
        Visit-specific parameters: Mid-transit time ($T_{\rm mid}$); planet-to-star radius ratio squared $(R_p/R_s)^2$; white noise ($s$); linear function slope ($v$); linear function constant ($c$); H$\alpha$ scale factor ($w_{\rm H\alpha}$); log GP amplitude ($\log_{10}(a_{\rm GP})$); log GP lengthscale ($\log_{10}(\lambda_{\rm GP})$). 
        Parameters fitted for each visit, then fixed to a weighted average (see Section~\ref{sec:lcfit}): Quadratic LD coefficients 1 and 2 ($q_1$, $q_2$) of each SOSS spectral order; impact parameter ($b$); normalized semi-major axis ($a/R_\star$). 
        Derived parameters: Planet-to-star radius ratio $(R_p/R_s)$ and its weighted average; absolute and normalized stellar density ($\rho_\star$, $\rho_\star/\rho_\odot$); normalized stellar radius ($R_\star/R_\odot$).
        $T_{\rm A21}$ is the predicted mid-transit time for each visit from \cite{agol_refining_2021}, in BJD - 2460000: -118.966820, 111.183168, 120.385481, 129.591582, and 148.003163 for Visits~1 to 5. The eccentricity is fixed to 0 and the argument of periapsis, to 90 degrees. 
        Parameters $\rho_\star$ and $\rho_\star/\rho_\odot$ are derived from $a/R_\star$ and the orbital period from \citet{ducrot_trappist-1_2020}. 
        Parameter $R_\star/R_\odot$ is derived from $\rho_\star/\rho_\odot$ and the stellar mass from \citet{ducrot_trappist-1_2020}. 
        For all posteriors with asymmetrical uncertainties, the reported values are based on the $16^{\rm th}$, $50^{\rm th}$, and $84^{\rm th}$ percentiles of the posterior distribution. For all weighted averages, the uncertainty is computed analytically via error propagation assuming all parameters are independent. 
        (F) indicates visits with a flare near ingress.
        }
        \label{tab:priors_posteriors_lcfit}
\end{deluxetable*}

\section{Retrieval and model comparison} \label{app:retrieval}

This appendix lists the priors of the planetary atmosphere and TLSE retrievals (Table~\ref{tab:retrieval_priors}). Table~\ref{tab:model_comparison} presents goodness of fit metrics, TLSE rejection significance, and surface pressure upper limits of a selection of atmosphere + TLSE models, compared across different combinations of visits for the \texttt{SOSSISSE} reduction.

\begin{deluxetable}{lc}[h]
        \tabletypesize{\scriptsize}
        \tablewidth{0pt} 
        \tablecaption{Priors of the planetary atmosphere and stellar contamination retrieval parameters.}
        \tablehead{
        \multicolumn{2}{c}{Planet} \\
        \colhead{Parameter} & \colhead{Prior}
        } 
        \startdata 
        Temperature, $T$ (K) & $\mathcal{U}(100, 300)$ \\
        Reference radius $R_{\rm p, ref}$ ($R_{\rm p}$) & $\mathcal{U}(0.85, 1.15)$ \\
        Log surface pressure, $\log_{10}{\left(P_{\rm surf}/{\rm bar}\right)}$ & $\mathcal{U}(-6, 2)$ \\
        Log volume mixing ratio, $\log_{10}{\left(X\right)}$ & $\mathcal{U}(-12, 0)$ \\
        Log Rayleigh-enhancement factor, $\log{a}$ & $\mathcal{U}(-4, 8)$ \\
        Scattering slope, $\gamma$ & $\mathcal{U}(-20, 0)$ \\
        \hline
        \multicolumn{2}{c}{Star} \\
        \colhead{Parameter} & \colhead{Prior} \\
        \hline
        Photosphere temperature, $T_{\rm phot}$ (K) & $\mathcal{N}(T_s, \sigma_{T_s}^2)$ \\
        Photosphere gravity, $\log_{10}{\left(g_{\rm phot}/{\rm cm\,s^{-2}}\right)}$ & $\mathcal{N}(5.2396, 0.01^2)$ \\
        \\
        \multicolumn{2}{c}{Single-heterogeneity fit} \\
        Heterogeneity covering fraction, $f_{\rm het}$ & $\mathcal{U}(0, 0.5)$ \\
        Heterogeneity temperature, $T_{\rm het}$ (K) & $\mathcal{U}(2300, 6000)$ \\
        Log heterogeneity gravity, $\log_{10}{\left(g_{\rm het}/{\rm cm\,s^{-2}}\right)}$ & $\mathcal{U}(3, 5.4)$ \\
        \\
        \multicolumn{2}{c}{Two-heterogeneity fit} \\
        Spot covering fraction, $f_{\rm het}$ & $\mathcal{U}(0, 0.5)$ \\
        Spot temperature, $T_{\rm het}$ (K) & $\mathcal{U}(2300, T_{\rm s} + 3\,\sigma_{T_{\rm s}})$ \\
        Facula covering fraction, $f_{\rm het}$ & $\mathcal{U}(0, 0.5)$ \\
        Facula temperature, $T_{\rm het}$ (K) & $\mathcal{U}(T_{\rm s} - 3\,\sigma_{T_{\rm s}}, 6000)$ \\
        \\
        \enddata
        \tablecomments{
        Not all parameters listed in this table were always fitted. For example, the (log) surface gravities were only fitted in a subset of retrievals to demonstrate that these additional parameters were not necessary. Elsewhere, all surface gravities were fixed to $\log{g}=5.2396$ \citep{agol_refining_2021}. 
        $R_p=1.045\,R_\oplus$, $T_{\rm s}=2566\,$K, and we adopted $\sigma_{T_{\rm s}}=50\,$K as a more conservative uncertainty, based on \citep{agol_refining_2021}.
        }
        \label{tab:retrieval_priors}
\end{deluxetable}

\begin{deluxetable*}{clcccccc}
        \tabletypesize{\scriptsize}
        \tablewidth{0pt} 
        \tablecaption{
        Log Bayesian evidence, reduced chi-squared, TLSE rejection confidence$^\dagger$, and 95\%-confidence upper limit on the surface pressure$^\ddagger$, with ``+h'' indicating that hazes are included in the model, compared across a selection of models and different combinations of visits for the \texttt{SOSSISSE} reduction.
        }
        \tablehead{
        \colhead{Model} & \colhead{Metric} & \colhead{Visit~2} & \colhead{Visit~3} & \colhead{Visit~5} & \colhead{Visits~2+3} & \colhead{Visits~2+3+5} & \colhead{Visits~1--5}
        } 
        \startdata 
        \\
        \multirow{2}{*}{Flat} & $\log{Z}$ & $363.03\pm0.10$ & $363.23\pm0.10$ & $349.18\pm0.09$ & $730.36\pm0.10$ & $1084.33\pm0.10$ & $1708.97\pm0.10$ \\
         & $\chi^2_{\rm red}$ & 1.00 & 0.59 & 0.51 & 0.77 & 0.67 & 0.82 \\
        \\
        \multirow{3}{*}{\shortstack{TLSE only}} & $\log{Z}$ & $359.77\pm0.13$ & $360.46\pm0.12$ & $346.47\pm0.12$ & $723.56\pm0.16$ & $1074.48\pm0.18$ & $1693.50\pm0.22$ \\
         & $\chi^2_{\rm red}$ & 1.03 & 0.62 & 0.54 & 0.79 & 0.70 & 0.83 \\
         & TLSE Rejection ($\sigma$) & 3.0 & 2.9 & 2.8 & 4.1 & 4.8 & 5.9 \\
        \\
        \multirow{5}{*}{\shortstack{H$_2$/He-dominated\\+multiple trace,\\with TLSE}} & $\log{Z}$ & $358.89\pm0.14$ & $359.34\pm0.13$ & $345.42\pm0.13$ & $722.31\pm0.17$ & $1073.08\pm0.19$ & $1700.66\pm0.22$ \\
        & $\chi^2_{\rm red}$ & 1.20 & 0.74 & 0.67 & 0.88 & 0.75 & 0.77 \\
        & TLSE Rejection ($\sigma$) & 2.9 & 3.0 & 2.9 & 4.0 & 4.8 & 5.9 \\
        & $P_{\rm surf}$ (bar) & $<2.39$ & $<0.0360$ & $<0.152$ & $<0.00447$ & $<0.00282$ & $<0.00162$ \\
        & $P_{\rm surf}$ (bar) +h & $^\star$ & $<0.0435$ & $<0.473$ & $<4.30$ & $<0.0175$ & $<12.8$ \\
        \\
        \multirow{5}{*}{\shortstack{N$_2$-dominated\\+multiple trace,\\with TLSE}} & $\log{Z}$ & $359.58\pm0.13$ & $360.32\pm0.12$ & $346.56\pm0.12$ & $723.27\pm0.16$ & $1073.86\pm0.19$ & $1693.70\pm0.24$ \\
        & $\chi^2_{\rm red}$ & 1.17 & 0.74 & 0.67 & 0.82 & 0.70 & 0.81 \\
        & TLSE Rejection ($\sigma$) & 3.1 & 2.9 & 2.9 & 4.1 & 5.0 & 6.3 \\
        & $P_{\rm surf}$ (bar) & $^\star$ & $^\star$ & $^\star$ & $^\star$ & $^\star$ & $^\star$ \\
        & $P_{\rm surf}$ (bar) +h & $^\star$ & $^\star$ & $^\star$ & $^\star$ & $^\star$ & $^\star$ \\
        \\
        \multirow{4}{*}{\shortstack{Pure CH$_4$,\\with TLSE}} & $\log{Z}$ & $357.99\pm0.14$ & $359.17\pm0.13$ & $345.48\pm0.12$ & $721.81\pm0.17$ & $1072.49\pm0.19$ & $1690.78\pm0.23$ \\
        & $\chi^2_{\rm red}$ & 1.09 & 0.65 & 0.57 & 0.81 & 0.71 & 0.84 \\
        & $P_{\rm surf}$ (bar) & $<0.326$ & $<3.49$ & $<8.45$ & $<0.00270$ & $<0.000411$ & $<0.000516$ \\
        & $P_{\rm surf}$ (bar) +h & $^\star$ & $<8.83$ & $<4.25$ & $^\star$ & $^\star$ & $^\star$ \\
        \\
        \multirow{4}{*}{\shortstack{Pure H$_2$O,\\with TLSE}} & $\log{Z}$ & $359.02\pm0.13$ & $360.27\pm0.13$ & $346.01\pm0.12$ & $722.83\pm0.16$ & $1073.86\pm0.18$ & $1692.97\pm0.22$ \\
        & $\chi^2_{\rm red}$ & 1.08 & 0.64 & 0.57 & 0.81 & 0.71 & 0.83 \\
        & $P_{\rm surf}$ (bar) & $<0.639$ & $<11.0$ & $<6.23$ & $<0.146$ & $<0.0422$ & $<4.87$ \\ 
        & $P_{\rm surf}$ (bar) +h & $^\star$ & $<9.73$ & $<12.5$ & $^\star$ & $^\star$ & $^\star$ \\ 
        \\
        \multirow{4}{*}{\shortstack{Pure NH$_3$,\\with TLSE}} & $\log{Z}$ & $360.69\pm0.13$ & $358.89\pm0.13$ & $346.32\pm0.12$ & $722.67\pm0.16$ & $1073.83\pm0.18$ & $1692.54\pm0.22$ \\
        & $\chi^2_{\rm red}$ & 1.01 & 0.66 & 0.56 & 0.81 & 0.71 & 0.84 \\
        & $P_{\rm surf}$ (bar) & $^\star$ & $<1.96$ & $^\star$ & $<10.7$ & $<3.37$ & $<1.71$ \\ 
        & $P_{\rm surf}$ (bar) +h & $^\star$ & $<1.51$ & $^\star$ & $^\star$ & $^\star$ & $^\star$ \\ 
        \\
        \enddata
        \tablecomments{
            {   
                $^\star$Unconstrained. 
                $^\dagger$TLSE rejection confidence levels are obtained by comparing the log Bayesian evidences of models with and without including the TLSE. For models that include an atmosphere (i.e., all models listed in this table except for the flat and TLSE-only models), the TLSE rejection confidence level is always computed for a haze-free atmosphere. 
                $^\ddagger$Upper limits on high-mean-molecular-mass atmospheres depend on the reduction pipeline.
            }
        }
        \label{tab:model_comparison}
\end{deluxetable*}

\section{Flare correction} \label{app:flarecorr}

In this appendix we detail the iterative approach we tested to correct for the flares in the spectral timeseries. While the resulting ``flare-corrected'' transit spectra were not used in the end because they still exhibited large-scale features unlikely to be planetary in origin, we present the method here as a possible starting point for future works.

The iterative flare correction can be summarized as follows:
\begin{enumerate}
    \item \label{enum:fluxcal} Absolute-flux-calibrate the spectral times series and mask all flare emission lines.
    \item Compute a median spectrum $m_i$ of all out-of-transit, out-of-flare spectra, where $i$ refers to the pixel number.
    \item \label{enum:fitspec} Fit each spectrum in the times series (including in-flare spectra and in-transit spectra) with the following model:
    \begin{equation}
        F_{i, j} = k_j m_i + a_j B_i(T_j) \,,
    \end{equation}
    where $F_{i, j}$ is the modeled flux at pixel $i$, at integration $j$, $k_j$ is a time-variable constant used to anchor the median spectrum, $a_j$ is a time-variable constant used to modulate the amplitude of the blackbody function $B_i$, which itself depends on a time-variable temperature $T_j$.
    \item Compute the relative correction factor at each pixel, at each integrations:
    \begin{equation}
        c_{i, j} = 1 - \frac{a^\star_j B_i(T^\star_j)}{\mathcal{F}_{i, j}} \,,
    \end{equation}
    where $a^\star_j$ and $T^\star_j$ are the best-fitting parameters from step~\ref{enum:fitspec}, and $\mathcal{F}_{i, j}$ is the observed flux from step~\ref{enum:fluxcal}.
    \item \label{enum:flarecorr} Apply the correction factors to the spectral time series that will be fed to the light curve fit.
    \item \label{enum:lcfit} Fit the corrected light curves from step~\ref{enum:flarecorr}.
    \item Subtract the best-fitting transit models from step~\ref{enum:lcfit} from the spectral time series from step~\ref{enum:fluxcal}.
    \item Go back to step~\ref{enum:fitspec}.
\end{enumerate}

This loop should be broken once the transit spectrum resulting from step~\ref{enum:lcfit} no longer varies significantly from one iteration to the next. To ensure the flare behaves in a somewhat physically plausible way, when fitting the spectra at step~\ref{enum:fitspec}, we applied a prior to $a_j$ and $T_j$ such that the two quantities are more likely to decrease monotonically in time after the flare peak.

We applied three iterations of this algorithm to Visits~1 and 4, but saw no improvement on the transit spectra. We suspect this is because when fitting the spectra in step~\ref{enum:fitspec}, there is a degeneracy between $k_j$ and $a_j$ such that $k_j$ can incorrectly absorb the flare signal, thus leaving the flare uncorrected when fitting the light curves at step~\ref{enum:lcfit}.

\section{Flare simulation} \label{app:flaresim}

\subsection{Forward model}

To explore the effect of flares on a transit spectrum, we simulated NIRISS-SOSS-like and NIRSpec-Prism-like spectral time series that include a planetary transit and a stellar flare. We started with a PHOENIX \citep{husser_new_2013} stellar spectrum at a temperature $T=2600\,$K, surface gravity $\log{g}=5.0$, iron abundance Fe/H$=0.0$, and alpha element abundance $\alpha/\mathrm{H}=0.0$. For SOSS-like simulations, we binned the spectrum down to the same wavelength grid we used to fit the SOSS light curves and made 121 copies of this stellar spectrum to produce a spectral time series similar to each SOSS visit of TRAPPIST-1\,f. For Prism-like observations, we binned the spectrum down to 50 bins equally spaced in wavelength and made 11\,117 copies of the stellar spectrum to mimic the TRAPPIST-1\,g observations of GO~2589 (Benneke et al., in review).

We then injected a planetary transit in each wavelength bin with the \texttt{batman} transit model \citep{kreidberg_batman_2015}, adopting all planetary and orbital parameters from \citet{gillon_seven_2017} and \citet{agol_refining_2021}, assuming a circular orbit, and using quadratic limb darkening coefficients computed with \texttt{ExoTiC-LD} \citep[][using a PHOENIX stellar model with $T=2600\,$K and $\log{g}=5.0$]{grant_exo-tic_2022,husser_new_2013}. In all simulations, we assumed that the planet has no atmosphere, such that the planet-to-star radius ratio is constant throughout all wavelength bins.

We then added a stellar flare signal that varies both in time and in wavelength. For the spectral dependence, we made the gross assumption that the flare behaves like a blackbody. 
The amplitude of this blackbody is computed with the temporal model from \citet{mendoza_llamaradas_2022}\footnote{\url{https://github.com/lupitatovar/Llamaradas-Estelares}}. The temporal model has three free parameters: the amplitude, the full width at half maximum (FWHM), and the flare peak time. In addition to this temporal model for the \textit{amplitude} of the blackbody, we assumed a different temporal dependence for the \textit{temperature} of the blackbody: before the flare peak, the blackbody has an initial temperature difference with respect to the photosphere of $\Delta T_0 > 0$; after the flare peak, the blackbody temperature decreases following an exponential law with the cooling time as a free parameter.

We then added random, Gaussian noise to the flux at each integration of each wavelength bin. For SOSS-like simulations, we used the time-medianed (relative) flux uncertainties of Visit~2 computed by \texttt{SOSSISSE}, scaled them by a factor of 0.1, binned them to the simulation wavelength grid, and multiplied them to the simulated flux to get the standard deviation ($\sigma_i$ in Equation~\ref{eq:flare_model}) of the Gaussian distribution from which the random noise was drawn. For Prism-like simulations, we used \texttt{PandExo} \citep{batalha_pandexo_2017} instead of real uncertainties, assuming 6 groups per integration (as in GTO 1201 and GO 2589). In both SOSS- and Prism-like simulations, the noise is not meant to be representative of the true, \textit{observed} noise; it is only meant to avoid numerical problems. Since the actual (observed) noise is likely not Gaussian, correlated, and larger than the values we used here, these simulations should be seen as ``best-case scenarios'' of transit observations contaminated by a stellar flare. 

The full forward model can be written as
\begin{equation}\label{eq:flare_model}
    F_{i, j} = F_{\star, i} \times \mathcal{T}_{i, j} + M_j \times B_i(T_j) + N_{i, j}(\sigma_i)\,,
\end{equation}
where $i$ is the wavelength bin number, $j$ is the integration number, $F_{i, j}$ is the modeled flux, $F_{\star, i}$ is the PHOENIX stellar photosphere flux \citep{husser_new_2013}, $\mathcal{T}_{i, j}$ is the \texttt{batman} transit model \citep{kreidberg_batman_2015}, $M_j$ is the \citet{mendoza_llamaradas_2022} model, $B_i$ is the blackbody, $T_j$ is the temperature of the blackbody, and $N_{i, j}$ is a random number drawn from a Gaussian distribution centered on 0 with standard deviation $\sigma_i$. The temperature of the blackbody is given by
\begin{equation}\label{eq:flare_model_temperature}
    T_j = \left\{\begin{array}{lr}
        T_\star + \Delta T_0, & \text{for } t_j < t_\mathrm{peak} \\
        T_\star + \Delta T_0\,e^{-(t_j - t_\mathrm{peak}) / t_\mathrm{cool}}, & \text{for } t_j \geq t_\mathrm{peak}
        \end{array}\right.\,,
\end{equation}
where $T_\star$ is the temperature of the stellar photosphere, $t_j$ is the time (e.g., in BJD) at integration $j$, $t_\mathrm{peak}$ is the flare peak time, and $t_\mathrm{cool}$ is the flare cooling time.

We then fitted the simulated light curves with \texttt{ExoTEP} \citep{benneke_spitzer_2017,benneke_water_2019,benneke_sub-neptune_2019} with the priors listed in Table~\ref{tab:priors_posteriors_lcfit}, but with several differences compared to the setup presented in Section~\ref{sec:lcfit}. We did not detrend against H$\alpha$ because the simulation does not produce emission lines. The limb darkening coefficients are fixed to their input values. In order to explore the impact of the flare on the transit spectrum (Section~\ref{app:flaresim_impact}), we looked at two scenarios: 1) we made minimal effort to correct for the flare to see the ``raw'' signature of the flare on the transit spectrum, and 2) we made maximal effort to correct for the flare to see the ``net'' impact of the flare. In scenario 1), we fitted the light curves with a transit model only (no systematics model), and masked all in-flare, out-of-transit integrations including post-transit integrations, i.e., the entire post-transit baseline is removed. In scenario 2), we fitted the light curves with a transit model, a linear function, and a SHO-kernel GP, and we masked in-flare, pre-transit integrations. Unlike in scenario 1), we did not mask post-transit integrations because we assumed that the GP would account for the remainder of the flare signal at those integrations. 

\begin{figure}
    \centering
    \includegraphics[width=.49\textwidth]{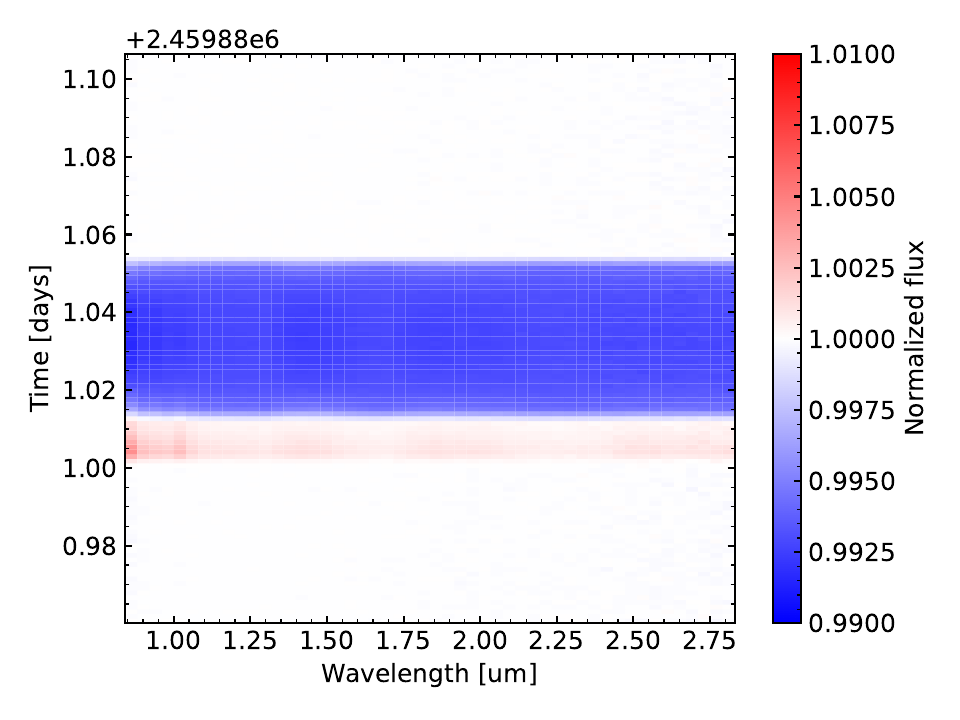}
    \includegraphics[width=.49\textwidth]{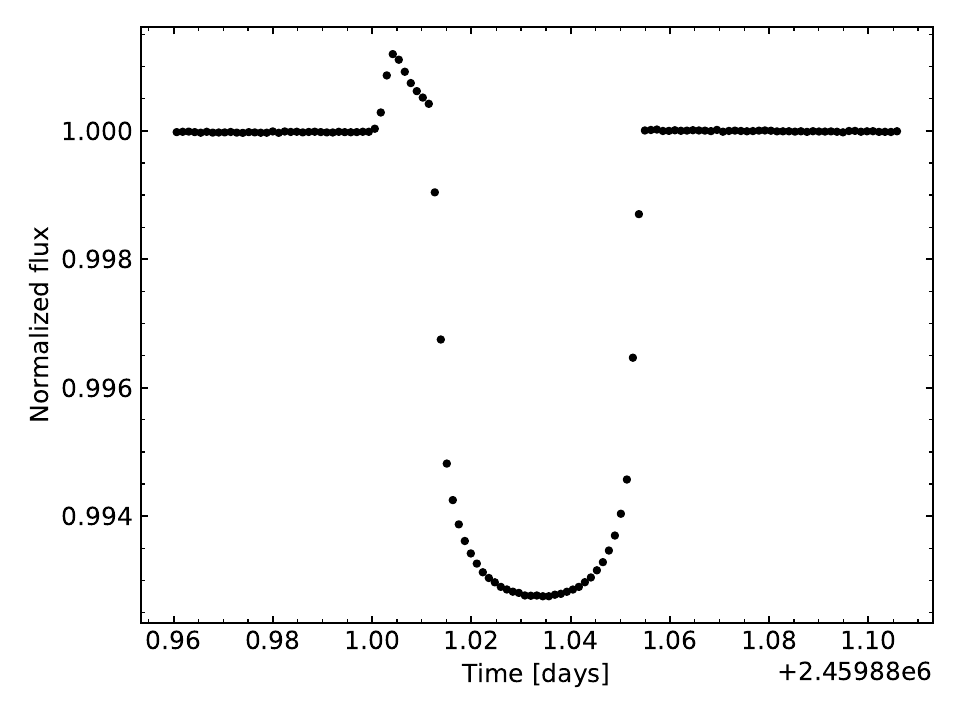}
    \caption{
    Example of flare simulation. Top: Spectral time series. Bottom: Broadband light curve. The flare has an amplitude of 0.001, FWHM of 10~minutes, peak time of 10~minutes prior to ingress, initial temperature difference of 1500\,K, and cooling time of 75~minutes. 
    }
    \label{fig:flaresim_2d_wlc}
\end{figure}

In summary, the flare simulation provides us with a transit spectrum given a set of stellar flare parameters. Figure~\ref{fig:flaresim_2d_wlc} shows an example of a simulated spectral time series with flare parameters chosen manually such that the broadband light curve would qualitatively match, as much as possible, the observed flare in Visit~1 of TRAPPIST-1\,f with SOSS in both spectral orders. 

\subsection{Raw and Net Impacts of Flares on Transit Spectra} \label{app:flaresim_impact}

\begin{figure*}
    \centering
    \includegraphics[width=\textwidth,trim={0 1em 0 0}]{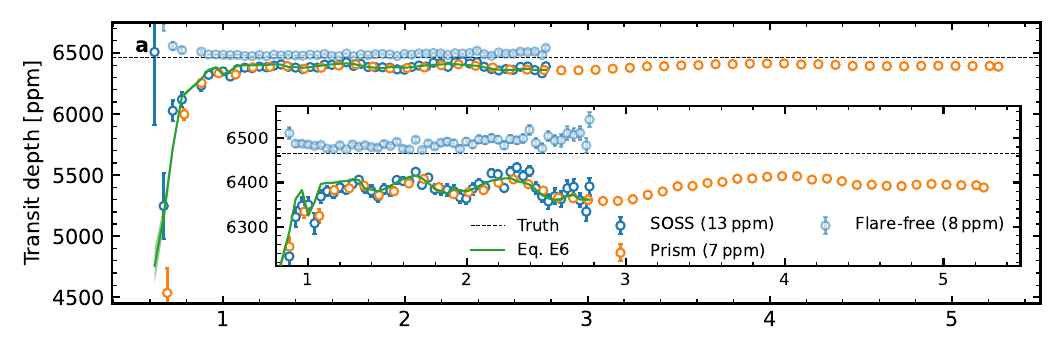}
    \includegraphics[width=\textwidth]{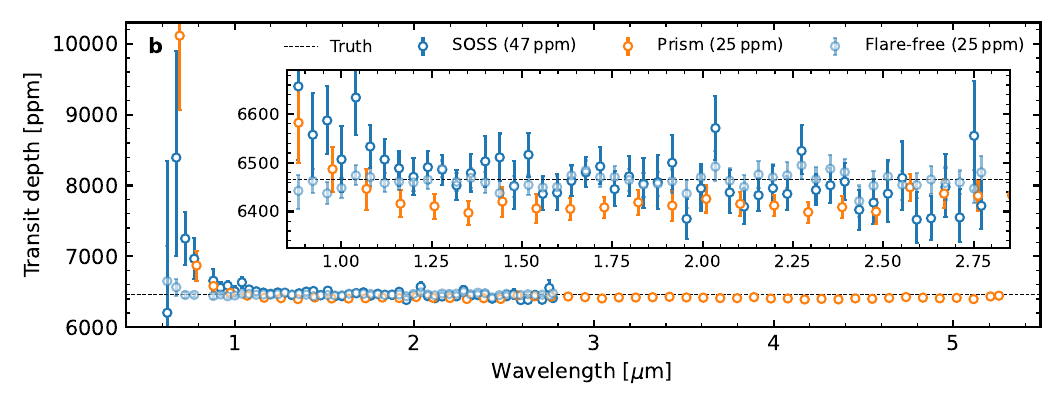}
    \caption{
    Examples of simulated, flare-contaminated transit spectra. 
    (a) Comparison between flare-contaminated transit spectra using NIRISS SOSS-like (solid blue) and NIRSpec Prism-like (orange) observational setups. Inset shows the spectra zoomed in. Light curves were fitted with a transit model only to show the ``raw'' effect of the flare. Green curve and shaded region show the median $\pm1\sigma$ from the fit of Equation~\ref{eq:flare_model_apparent}.
    (b) Same as panel (a), but light curves were fitted with a transit model, a linear function, and a simple-harmonic-oscillator-kernel Gaussian process to show the ``net'' effect of the flare. Inset shows the spectra zoomed in. 
    In both panels, the transparent blue spectrum is a flare-free simulation using the SOSS-like observational setup, the black dotted line is the true (input) squared planet-to-star radius ratio, and numbers in parentheses in the legends give the median transit depth error bar.
    }
    \label{fig:flaresim_tspec}
\end{figure*}

The blue and orange spectra in Figure~\ref{fig:flaresim_tspec}a were obtained from SOSS-like and Prism-like simulated spectral times series, respectively, with the same flare parameters as in Figure~\ref{fig:flaresim_2d_wlc}, with no planetary atmosphere, and by fitting the light curves with a transit model only to show the ``raw'' impact of the flare. Both spectra have a slope at wavelengths bluewards of about 1.1\,$\mu$m as well as bumps resembling inverted water absorption bands, similar to the transit spectrum of unocculted faculae. We note that the SOSS-like spectrum is very different from the spectra we obtained by fitting the light curves of Visits~1 and 4 (the two visits with a flare near the ingress) with minimum effort to account for the flare, indicating that the simulation is probably too simple to capture the true complexity of TRAPPIST-1 flares and/or TRAPPIST-1 activity and/or other systematics (e.g., from the data reduction). 

To help understand why we see faculae-like features in the simulated, flare-contaminated transit spectra, we can analytically derive an expression for the (wavelength-dependent) apparent transit depth $d_{i, \rm app}$ using Equation~\ref{eq:flare_model} without the noise term. We first define $d_{i, \rm app}$ as follows:
\begin{equation}\label{eq:flare_model_apparent_0}
    d_{i, \rm app} = \frac{F_{i, \rm out} - F_{i, \rm in}}{F_{i, \rm out}}\,.
\end{equation}
$F_{i,\rm out}$ is the (wavelength-dependent) out-of-transit, out-of-flare flux, i.e., the flux at an integration $j_{\rm out}$ such that $M_{j_{\rm out}}$ is zero. $F_{i, \rm in}$ is the mid-transit flux, which we assume to be at integration $j_{\rm mid}$, and we assume that $M_{j_{\rm mid}}$ is not zero. By substituting $F_{i, \rm out}$ with $F_{\star, i}$ and $F_{i, \rm in}$ with $F_{\star, i} \times \mathcal{T}_{i, j_{\rm mid}} + M_{j_{\rm mid}} \times B_i(T_{j_{\rm mid}})$ in Equation~\ref{eq:flare_model_apparent_0}, we obtain 
\begin{equation} \label{eq:flare_model_apparent}
    d_{i, \rm app} = d_{i, 0} - \frac{M_{j_{\rm mid}} B_i(T_{j_{\rm mid}})}{F_{\star, i}} \,,
\end{equation}
where $d_{i, 0} = 1 - \mathcal{T}_{i, j_{\rm mid}}$ is the apparent transit depth in the absence of flares (wavelength-independent if no atmosphere was injected). The first thing to notice is that $d_{i, \rm app} < d_{i, 0}$, as seen in panel a of Figure~\ref{fig:flaresim_tspec}, which translates the fact that the flare dilutes the transit. The second thing to notice is that $M_{j_{\rm mid}}$ is wavelength-independent and $B_i(T_{j_{\rm mid}})$ is essentially a downward slope redwards of $\sim1\,\mu$m for temperatures $\sim$4000\,K, the approximate temperature of the flare at mid-transit in the simulation. The apparent transit depth should thus resemble the apparent transit depth in the absence of flares ($d_{i, 0}$), diluted by a slope ($M_{j_{\rm mid}} B_i(T_{j_{\rm mid}})$) which is modulated by the inverse of the quiet photosphere spectrum ($F_{\star, i}$). The inverted absorption-band-like features in panel a of Figure~\ref{fig:flaresim_tspec} are thus essentially the (inverted) quiet photosphere spectrum. The green curve and shaded region show the median $\pm1\sigma$ model from fitting Equation~\ref{eq:flare_model_apparent} to the SOSS-like transit spectrum with $M_{j_{\rm mid}}$ and $T_{j_{\rm mid}}$ as free parameters.

In panel b of Figure~\ref{fig:flaresim_tspec}, the blue and orange spectra correspond once again to SOSS- and Prism-like simulations, respectively, but the light curves were fitted with a transit model, a linear function, and a SHO-kernel GP to show the ``net'' impact of the flare. The two spectra generally agree again, and we recover the flat spectrum at the expected value for wavelengths longer than about 1\,$\mu$m. However the systematics treatments are unable to capture the flare signal at shorter wavelengths, as demonstrated by the increase in transit depths bluewards of 1\,$\mu$m. This increase is somewhat similar to the observed increase in Visit~1, but the simulations do not reproduce the observed, downward slope redwards of 1\,$\mu$m and the increase at the reddest wavelengths. We also note that the median transit depth uncertainty, provided in the legend, is larger than in panel a due to degeneracies between $(R_p/R_\star)^2$, the GP hyperparameters, and the linear function parameters. 

In summary, panel b of Figure~\ref{fig:flaresim_tspec} shows that in a ``best-case scenario'' of a flare occurring shortly before the transit ingress (Gaussian, uncorrelated noise; orbital parameters and limb darkening coefficients known with 100\% certainty; no planetary atmosphere; no TLSE; blackbody flare model), fitting the light curves with a linear function and a SHO-kernel GP returns the expected flat spectrum at wavelengths redder than about 1\,$\mu$m. Outside of this restricted space, it is unclear how unreliable a transit spectrum can get, further supporting the need for a better understanding of stellar flares.


\bibliography{references,extra}{}
\bibliographystyle{aasjournalv7}



\end{document}